\documentclass[]{iopart}

\expandafter\let\csname equation*\endcsname\relax
\expandafter\let\csname endequation*\endcsname\relax
\usepackage{graphicx}
\usepackage{epstopdf}
\usepackage{cite}
\usepackage[perpage,marginal]{footmisc}
\usepackage{etoolbox}
\usepackage{marvosym}
\usepackage{upgreek}
\usepackage{pdflscape}
\usepackage[unicode]{hyperref}
\usepackage[all]{hypcap}
\usepackage{amsmath,amsfonts,amssymb}
\usepackage{color}
\usepackage[usenames,dvipsnames]{xcolor}
\usepackage{bm}
\usepackage{longtable}
\usepackage[T1]{fontenc}
\usepackage[utf8]{inputenc}
\usepackage{verbatim}
\usepackage{pifont}
\usepackage{yfonts}
\usepackage{url}
\usepackage{multirow}
\usepackage{tensor}
\usepackage[normalem]{ulem}
\usepackage{mathtools}
\makeatletter
\def\url@leostyle{
  \@ifundefined{selectfont}{\def\UrlFont{\sf}}{\def\UrlFont{\small\ttfamily}}}
\makeatother
\usepackage[normalem]{ulem}
\usepackage{fancyhdr}
\makeatletter
\renewcommand\footnoterule{
  \kern-3\p@
  \hrule\@width2.5cm
  \kern2.6\p@}
\makeatother
\numberwithin{equation}{section}
\numberwithin{figure}{section}
\begin{document}

\pagestyle{fancy}
\lhead{Stellar collapse in scalar-tensor theories}
\chead{}
\rhead{\thepage}
\lfoot{}
\cfoot{}
\rfoot{}

\begin{center}
\title{Numerical simulations of stellar collapse in scalar-tensor theories of gravity}
\end{center}

\author{
Davide~Gerosa$^{1}$,
Ulrich~Sperhake$^{1,2,3}$
and\\
Christian D. Ott$^{2,4}$
}

\address{$^{1}$~Department of Applied Mathematics and Theoretical Physics, Centre for Mathematical Sciences, University of Cambridge, Wilberforce Road, Cambridge CB3 0WA, UK}
\address{$^{2}$~TAPIR, Walter Burke Institute for Theoretical Physics, California Institute of Technology, Pasadena, CA 91125, USA}
\address{$^{3}$~Department of Physics and Astronomy,
The University of Mississippi, University, MS 38677-1848, USA}
\address{$^{4}$~Yukawa Institute for Theoretical Physics, Kyoto University, Kyoto, Japan}

\ead{d.gerosa@damtp.cam.ac.uk}

\begin{abstract}
We present numerical-relativity simulations of spherically symmetric
core collapse and
compact-object
formation in scalar-tensor theories of gravity.
The additional scalar degree of freedom introduces a propagating
monopole gravitational-wave mode.  Detection of monopole scalar waves
with current and future gravitational-wave experiments may constitute
smoking gun evidence
for
strong-field modifications of General Relativity. We collapse both
polytropic and more realistic pre-supernova profiles using a
high-resolution shock-capturing scheme and an approximate prescription
for the nuclear equation of state. The most promising sources of
scalar radiation are protoneutron stars collapsing to black holes. In
case of a Galactic core collapse event forming a black hole, Advanced
LIGO may be able to place independent constraints on the parameters of
the theory at a level comparable to current Solar-System and
binary-pulsar measurements. In the region of the parameter space
admitting spontaneously scalarised stars, transition to configurations
with prominent scalar hair before black-hole formation further enhances the
emitted signal.  Although a more realistic treatment of the
microphysics is necessary to fully investigate the occurrence of
spontaneous scalarisation of neutron star remnants, we speculate that
formation of such objects could constrain the parameters of the
theory beyond the current bounds obtained with Solar-System and
binary-pulsar experiments.
\end{abstract}

\mathindent = 0pc


\section{Introduction}
\label{sec:intro}

General Relativity (GR) is currently assumed to be the standard
theory of gravity, and has so far passed all experimental
tests with flying colours
\cite{2015CQGra..32x3001B,2016arXiv160203841T,2014LRR....17....4W,2008LRR....11....9P}.
Theoretical and observational evidence, however, suggests that
some modifications of GR may be inevitable.
Cosmological
and astrophysical observations
require most of the energy content of the Universe to
be present in the form of dark energy and dark matter \cite{2005PhR...405..279B,2006IJMPD..15.1753C,2015Sci...347.1100S}.
On theoretical grounds, GR is expected to represent
the low-energy limit of a more fundamental (quantum) theory \cite{2004LRR.....7....5B}.
Presently considered candidates for such theories predict
modifications of GR at higher energies which also provide
means to circumvent the formation of mathematical singularities
inevitable in GR \cite{1970RSPSA.314..529H}.

Attempts to generalise GR in these directions often involve
additional fields that mediate the gravitational interaction together
with the spacetime metric. The simplest class of such models
is that of scalar-tensor (ST) theories,
where one scalar field is included in the gravitational
sector of the action. Ever since the
pioneering work of Jordan, Fierz, Brans, and Dicke
\cite{1959ZPhy..157..112J,Fierz:1956zz,1961PhRv..124..925B}, ST
theories have received a great deal of attention, both from a theoretical and a phenomenological point of view (see e.g. \cite{1992CQGra...9.2093D,1997PThPS.128..335C,2003sttg.book.....F,2015LNP...892....3S,2004cstg.book.....F} and references therein).
This class of theories is {\it simple enough} to allow for detailed
predictions to be worked out, but also {\it complicated enough} to
introduce a richer phenomenology leading to potentially observable
deviations from GR. ST theories make 
predictions in the largely untested strong-field regime, while
remaining compatible with the weak-field constraints imposed on GR by
Solar System experiments (cf. Sec.~\ref{coupfun} below).

Black hole (BH) spacetimes might at first glance appear to represent an ideal
testing ground for strong-field effects.
 The classical no-hair theorems,
first proven  for Brans-Dicke theory  
\cite{1972CMaPh..25..167H,1971ApJ...166L..35T,Chase:1970} and later extended to
a wider range of ST theories (see \cite{1997siad.conf..216B,2015IJMPD..2442014H}
for reviews), however, strongly constrain the potential for
deviations of BH spacetimes in ST theory from their GR counterparts.
At leading post-Newtonian (PN) order, for example, the dynamics of a BH binary
in Brans-Dicke theory is indistinguishable from the GR case
\cite{1989ApJ...346..366W}. Indeed, considering the ST field equations
given below as Eqs.~(\ref{jordanevol})-(\ref{jordanevolphi}),
one immediately sees that vacuum solutions of GR are also solutions
to the ST equations with a constant scalar field.
Non-trivial BH dynamics can still be obtained by relaxing some of the
fundamental ingredients of the no-hair theorems as for example
a non-vanishing potential term \cite{2012CQGra..29w2002H} or
non-asymptotic flatness \cite{2013PhRvD..87l4020B}. Due to the additional
coupling introduced by the energy momentum tensor in the
ST equations, however, compact matter sources such as neutron stars (NSs) and collapsing protoneutron stars forming BHs
appear to be more promising objects for exploring observational
signatures of ST theories. 

Guided by this expectation, we shall focus in this paper on the
formation of compact objects through gravitational collapse.
Gravitational collapse is the expected evolutionary endpoint of stars
of zero-age main sequence  
(ZAMS)
mass of $10
M_{\odot} \lesssim M \lesssim 130 M_{\odot}$
\cite{2011ApJ...730...70O,2015ApJ...799..190C,2016ApJ...821...38S}. After
exhausting their available fuel, the star's central core (mostly made
of iron group nuclei) collapses under the strength of gravity as it
exceeds its effective Chandrasekhar mass
\cite{1990RvMP...62..801B}. Collapse proceeds until mass densities
become comparable to those of nuclear matter. Thereafter, the
increasingly repulsive character of the nuclear interactions leads to
core bounce, which results in an outgoing hydrodynamic shock.  The
outgoing shock soon stalls because of dissociation of nuclei and
neutrino emission
in the
post-shock region, and must be revived to successfully drive a
supernova (SN) explosion \cite{1990RvMP...62..801B}. The physical
mechanism responsible for the shock revival is still a topic of active
research (see e.g. \cite{2007PhR...442...38J} and references
therein). Multi-dimensional fluid instabilities and neutrino
interactions are generally believed to play a crucial role in driving
most SN explosions with the possible exception of hyper-energetic ones
\cite{2007ApJ...664..416B,2012ApJ...757...69U}.  One single
core-collapse SN provides photon luminosities comparable to those of
an entire galaxy and outshines all stars in the Universe in
neutrinos. 
If the explosion is successful, a NS is left behind. If the explosion fails or is
  very weak, continued accretion will push the central NS over its
  maximum mass of $2-3\,M_\odot$ and lead to the formation of a
  BH. The details of BH formation depend on the structure of the
  progenitor star and on the nuclear equation of state (EOS)
  \cite{2011ApJ...730...70O}. 

ST theories may play a crucial role in this picture of NS and BH
formation. A peculiar non-linear effect called ``spontaneous
scalarisation'' \cite{1993PhRvL..70.2220D,1996PhRvD..54.1474D}
-- somewhat similar to spontaneous magnetisation in ferromagnets --
represents a particularly strong form of non-trivial scalar-field dynamics
leading to additional branches of stationary NS families (see also \cite{2013PhRvD..87h1506B,2014PhRvD..89d4024P,2014PhRvD..89h4005S} for dynamical scalarisation in binary NS systems). Moreover, ST theories provide a new channel for emission of
gravitational waves (GWs) in stellar collapse.
Whereas in GR conservation of mass and momentum
exclude monopole and dipole radiation, monopole waves
are permitted in ST theories in the form of scalar radiation,
the so-called {\it breathing mode}.
Detection of this breathing mode generated by a galactic SN
would constitute smoking-gun evidence for a deviation from GR in the strong-field regime.
Such tests of GR represent a major scientific goal
\cite{2016arXiv160203841T}
of the new era of GW astronomy initiated with the
recent breakthrough detection of GW150914 \cite{2016PhRvL.116f1102A},
and thus add to the enormous scientific potential
of exploring the physics of stellar collapse with GWs
(see Ref.~\cite{2009CQGra..26f3001O} for a comprehensive
review on the topic).

The impact of ST theories on the equilibrium structure of NSs has been
extensively studied in the literature (see, e.g.,
\cite{1993PhRvL..70.2220D,2013PhRvD..88h4060D,2015PhRvD..91f4024M,2015CQGra..32n5008S,2015CQGra..32t4001H,2016PhRvD..93d4009P}).
Surprisingly few studies, however, have been devoted to their
formation processes.
Following pioneering numerical relativity simulations in Brans-Dicke theory \cite{1973PThPh..49.1195M}, early studies have been devoted to dust-fluid collapse \cite{1995PhRvD..51.4208S,1995PhRvD..51.4236S,1994PhRvD..50.7304S,1997PhRvD..55.2024H}. The collapse of NSs into BHs \cite{1998PhRvD..57.4789N} and the transition between different static NS branches \cite{1998PhRvD..58f4019N} was first addressed by Novak using pseudo-spectral methods. 
To the best of our knowledge, the only published simulations
of NS formation in ST theories have been presented by Novak and Ib{\'a}{\~n}ez
in Ref.~\cite{2000ApJ...533..392N}, who combined
  pseudo-spectral techniques and high-resolution shock-capturing to
  study core collapse.   The only other study we are aware of is
Ref.~\cite{dimmthesis}, which numerically models the collapse of
spherically symmetric fluids with 
 a  $\Gamma$-law
  EOS in Brans-Dicke theory and finds the monopole
radiation to dominate at frequencies near the GW detectors' maximum
sensitivity regime $f\sim100~{\rm Hz}$, independently of the
Brans-Dicke coupling parameter.  {The systematic exploration of GW
  emission from core collapse in ST theories thus represents a largely
  uncharted area in SN research. The dawning age of observational GW
  physics makes the filling of this gap a timely task, the first step
  of which is the main goal of this paper.}

{For this purpose, we have extended} the open-source code
\textsc{gr1d} of O'Connor and Ott \cite{2010CQGra..27k4103O} to ST
theory and performed numerical simulations {of NS and BH formation
  following core collapse} to address the detectability of the
monopole GWs with Advanced LIGO \cite{2015CQGra..32g4001T,2016arXiv160203838T} and the
proposed
 Einstein Telescope
\cite{2010CQGra..27s4002P}. We tackle the following questions.
\begin{itemize}
\item Are non-trivial scalar-field profiles {and correspondingly
large amplitudes in the scalar radiation} naturally triggered in
compact remnants following stellar collapse?
\item Can future GW observations of core collapse provide smoking gun
  {evidence} 
 for deviations from GR
      in the framework of ST theories?
\end{itemize}

This paper is organised as follows. The action and the evolution
equations of the theory are presented in Sec.~\ref{eveqsec}.
 Additional physical ingredients entering our simulations are given
in Sec.~\ref{physicalsetup}. Our numerical procedure is  described
in Sec.~\ref{numimpl}. We  present our results on core collapse
dynamics and monopole GW emission in Sec.~\ref{results}. We 
summarise our findings in Sec.~\ref{conclusions}.  Supporting material is provided online at Ref.~\cite{DGwebsite}.
Throughout the
paper, we generally use geometrical units $c=G=1$, but occasionally
restore factors of $G$ for clarity of presentation.

\section{Evolution equations}
\label{eveqsec}

In this Section, we first review different ways to formulate ST theories
and then arrive at the equations for the metric, scalar field, and matter
sector in general covariant form (Sec.~\ref{formulations}).
Next, we derive the hydrodynamic equations for the matter sources, the metric and scalar field for the specific
case of radial-gauge, polar-slicing coordinates (Sec.~\ref{hydrospherical}).

\subsection{A tale of two formulations}
\label{formulations}

In ST theories, gravity is mediated by the spacetime metric $g_{\mu
\nu}$ and an additional scalar field $\phi$. The most general action
which (i) involves a single scalar field coupled non-minimally to
the metric, (ii) is invariant under space-time diffeomorphisms,
(iii) is at most quadratic in derivatives of the field, and (iv) satisfies the
weak equivalence principle can be written in the form
\cite{1992CQGra...9.2093D,2015CQGra..32x3001B,2006CQGra..23.4719S}
\begin{equation}
  S = \int dx^4 \sqrt{-g} \left[ \frac{F(\phi)}{16\pi G} R
      - \frac{1}{2}g^{\mu \nu} (\partial_{\mu} \phi)(\partial_{\nu}\phi)
      - V(\phi) \right] + S_m(\psi_m,g_{\mu \nu})\,.
\label{Jaction}
\end{equation}
Here, $d^4x$ is the standard coordinate volume element,
$R$ is the Ricci scalar built from $g_{\mu \nu}$,
$g=\det g_{\mu \nu}$ and the
symbol $\psi_m$ collectively denotes all non-gravitational fields.
The theory has only two free functions of the scalar field: the
potential $V=V(\phi)$ and the coupling function $F=F(\phi)$\footnote{Another common notation for the coupling function is $A=F^{-1/2}$ (see, e.g., Refs. \cite{1992CQGra...9.2093D,1993PhRvL..70.2220D,1996PhRvD..54.1474D}) .}. If the
potential $V$ is a slowly varying function of $\phi$ -- as expected
on cosmological grounds, see \cite{2007arXiv0704.0749D} --  it causes
negligible effects on the propagation of $\phi$ on stellar scales.
For simplicity, we thus set $V=0$ throughout this paper;
GR is then recovered for $F=1$. Details on the choice of the coupling
function $F$ are postponed to Sec.~\ref{coupfun}.

The weak equivalence principle -- which has been verified
experimentally to very high precision \cite{2014LRR....17....4W} -- is
guaranteed to hold as long as the matter part of the action $S_m$ does
not couple to the scalar field, and its motion is therefore governed
by the geodesics of the metric $g_{\mu\nu}$. In this formulation, the
scalar field does not interact with ordinary matter directly, but
influences the motion of particles exclusively through its coupling
with the spacetime metric.

The theory described by the action (\ref{Jaction}) is said to be
formulated in the \emph{Jordan frame} \cite{1959ZPhy..157..112J}.
Probably the most famous case of a ST theory, though by now
severely constrained by solar-system tests \cite{2003Natur.425..374B}, is
Brans-Dicke theory \cite{1961PhRv..124..925B}: the specific theory
obtained by setting $F(\phi)= 2\pi \phi^2/ \omega_{\rm BD}$ where
$\omega_{\rm BD}$ is constant \cite{2006CQGra..23.4719S}.

Alternatively to the
above Jordan-frame description, ST theories can also be formulated
in the so-called \emph{Einstein frame}. Here, one considers the conformal
transformation
\begin{align}
\bar g_{\mu\nu} = F g_{\mu\nu}\,,\label{eq:conformalTrafo}
\end{align}
and the action of Eq.~(\ref{Jaction}) becomes
\begin{equation}
  S = \frac{1}{16\pi G} \int dx^4 \sqrt{-\bar{g}} \left[
      \bar{R} - 2\bar{g}^{\mu \nu} (\partial_{\mu} \varphi)
      (\partial_{\nu} \varphi) \right]
      +S_m[\psi_m,\bar{g}_{\mu \nu}/F]\,.
  \label{Eaction}
\end{equation}
The Ricci scalar $\bar{R}$ is now built from the Einstein metric
$\bar g_{\mu \nu}$
and $\varphi$ is a redefinition of the scalar field $\phi$ through
\cite{2006CQGra..23.4719S,2008PhRvD..77j4010S},
\begin{equation}
  \frac{\partial \varphi}{\partial \phi} =  \sqrt{\frac{3}{4} \frac{F_{,\phi}{}^2}{F^2} + \frac{4\pi G}{F}}\,.
  \label{fieldredefinition}
\end{equation}
The key advantage resulting from this conformal transformation is
a minimal coupling between the conformal metric and the scalar
field, evident at the level of the action. The fact that such a
redefinition of the theory exists has an important consequence
for attempts to constrain the theory's parameters through
observations of compact objects:
BHs are less suitable to obtain such constraints
because the action (\ref{Eaction}) in vacuum ($S_m=0$)
reduces to the Einstein-Hilbert action of GR with a minimally coupled
scalar field.
In the action as well as the field equations given further below, it
is evident that matter sources represent an additional and more
straightforward channel  
to couple the metric and scalar sectors.

The equations of motion in the Jordan frame can be obtained by varying
the action (\ref{Jaction}) with respect to the spacetime
metric $g_{\mu\nu}$ and the scalar field $\phi$:
\begin{align}
  G_{\mu\nu} &= \frac{8\pi}{F} \left(
       T_{\mu\nu}^F + T_{\mu\nu}^{\phi}
       + T_{\mu\nu} \right)\,, \label{jordanevol} \\
  T_{\mu\nu}^F &= \frac{1}{8\pi} ( \nabla_{\mu}
      \nabla_{\nu}F - g_{\mu\nu} \nabla^{\rho}\nabla_{\rho}F)\,,
      \label{jordanevol2}  \\
  T_{\mu\nu}^{\phi} &= \partial_{\mu} \phi
       \partial_{\nu} \phi -\frac{1}{2}  g_{\mu\nu}
     \partial^{\nu} \phi
       \partial_{\nu} \phi \, , \label{jordanevol3} \\
  \nabla^{\rho}\nabla_{\rho} \phi &= -\frac{1}{16\pi} F_{,\phi}R\,.
       \label{jordanevolphi}
\end{align}
Combining the Bianchi identities with the field equations can be shown
to imply that the matter part of the energy momentum tensor,
\begin{align}
  T_{\mu\nu} = \frac{2}{\sqrt{-g}} \frac{\delta S_m}{\delta g_{\mu\nu}}\,,
\end{align}
is conserved on its own, i.e.
\begin{equation}
  \nabla_{\mu} T^{\mu \nu} = 0\,. \label{eq:nablaT}
\end{equation}
This feature  makes the Jordan frame particularly suitable for
studying stellar collapse: the matter equations, which are
expected to develop shocks, do not need to be modified from their
GR counterparts (cf. Sec.~\ref{mattereq}). The drawback of this
choice is that the scalar field is not minimally coupled to the
metric, i.e.~the Hilbert term in the action (\ref{Jaction})
acquires a $\phi$-dependent factor. This factor $F(\phi)$
leads to the term $T_{\mu \nu}^F$ on the right-hand side
of Eq.~(\ref{jordanevol}) additionally to the
minimally coupling term  $T_{\mu \nu}^{\phi}$
and the standard matter sources $T_{\mu \nu}$.

\subsection{Equation of motions}
\label{hydrospherical}

We now restrict the equations of motion to spherical symmetry in
radial-gauge, polar-slicing  coordinates \cite{1997stgr.proc..289R}.
The line element in the Jordan frame is
\begin{equation}
  ds^2 = g_{\mu\nu}dx^{\mu}dx^{\nu}  = -\alpha^2 dt^2 + X^2dr^2
        + \frac{r^2}{F} d\Omega^{2}\,,
        \label{lineelement}
\end{equation}
where the metric functions $\alpha=\alpha(t,r)$ and $X=X(t,r)$ can
be more conveniently rewritten in terms of the metric potential,
\begin{align}
  \Phi = \ln(\sqrt F \alpha)\,,\label{eq:Phi}
\end{align}
and the enclosed mass,
\begin{align}
  m = \frac{r}{2} \left(1- \frac{1}{FX^2} \right)\,. \label{eq:m}
\end{align}
Note that in Eq.~(\ref{lineelement}) we multiplied the angular part
of the metric $d\Omega^{2}$ by a factor $1/F$,  thus effectively
imposing the radial gauge in the Einstein frame. In this formulation,
the (Jordan-frame) areal radius is given by $r/\sqrt{F}$. This
choice allows for comparisons with
Refs.~\cite{1998PhRvD..57.4789N,1998PhRvD..58f4019N,2000ApJ...533..392N},
where the analysis is entirely carried out in the Einstein frame.
Likewise, $\Phi$ and $m$ are Einstein-frame variables and their definition
in terms of the Jordan metric components in Eqs.~(\ref{eq:Phi}),
(\ref{eq:m}) acquires factors of $F$.

Following \cite{2010CQGra..27k4103O}, we assume ideal hydrodynamics
as described by an energy-momentum tensor of the form
\begin{align}
  T_{\alpha \beta} &= \rho h u_{\alpha} u_{\beta} + P g_{\alpha \beta}\,,
\end{align}
and the matter current density
\begin{align}
    J^{\alpha} &= \rho u^{\alpha}\,.
\end{align}
Here $\rho$ is the baryonic density, $P$ is the fluid pressure,
$h$ is the specific enthalpy (which is related to the specific
internal energy $\epsilon$ and the pressure $P$ by $h=1+\epsilon+P/\rho$), and $u^\mu$
is the 4-velocity of the fluid. Spherical symmetry implies
\begin{equation}
  u^{\mu} = \frac{1}{\sqrt{1-v^2}}
      \left[
        \frac{1}{\alpha},~\frac{v}{X},~0,~0
      \right],
\end{equation}
where $v=v(t,r)$.

The equations of motion can be reformulated in flux conservative form
using \emph{conserved} variables and thus become amenable to
a numerical treatment using
high-resolution shock-capturing schemes
\cite{2000PhRvD..61d4011F,2008LRR....11....7F}. These conserved
variables  $D$, $S^r$ and $\tau$ are related to the to the \emph{primitive}
variables $\rho$, $\epsilon$, $v$ and $P$ by
\begin{align}
  D &= \frac{\rho X}{F\sqrt{F} \sqrt{1-v^2}}\,, \label{defD} \\
  S^r &= \frac{\rho h v}{F^2(1-v^2)}\,, \label{defS} \\
  \tau  &= \frac{\rho h}{F^2(1-v^2)} - \frac{P}{F^2}-{D} \label{deftau}\,.
\end{align}
The definitions above generalise Eq.~(8) in Ref.~\cite{2010CQGra..27k4103O}
to ST theory.
We take advantage of the
Einstein-frame scalar-field redefinition $\phi \to \varphi$ of
Eq.~(\ref{fieldredefinition}) because it simplifies the
wave equation (\ref{jordanevolphi}). Moreover, the space of ST
theories and the weak-field experimental constraints are traditionally
described in terms of  $\varphi$ (cf.\ Sec.~\ref{coupfun}).
Following
Refs.~\cite{1998PhRvD..57.4789N,1998PhRvD..58f4019N,2000ApJ...533..392N},
we introduce auxiliary variables for
the derivatives of the scalar field defined by
\begin{align}
  \eta = \frac{\partial_r \varphi}{X}\,, \qquad
  \psi = \frac{\partial_t \varphi}{\alpha} . \label{eq:psi}
\end{align}

\subsubsection{Metric equations.}
\label{metriceq}
The evolution equations (\ref{jordanevol}--\ref{jordanevol3}) for
the metric potential $\Phi$ and the mass function $m$ expressed
in terms of the conserved variables read
\begin{align}
  \partial_r \Phi =& X^2F
      \left[
        \frac{m}{r^2} +4\pi r \left( {S}^r v + \frac{P}{F^2} \right)
        +\frac{r}{2F} (\eta^2 + \psi^2)
      \right], \label{eq:Phir} \\
  \partial_r m =& 4\pi r^2 ({\tau}+{D})
      + \frac{r^2}{2F}(\eta^2 + \psi^2)\,, \label{eq:mr} \\
  \partial_t m =& r^2\frac{\alpha}{X}
      \left( \frac{1}{F} \eta \psi - 4\pi {S}^r \right). \label{eq:mt}
\end{align}
These equations are not independent;
the last equation for $\partial_t m$ directly follows from
the other two combined with the conservation of the energy momentum tensor
(\ref{eq:nablaT}). For convenience, we follow standard practice and compute
the metric functions using the constraints (\ref{eq:Phir}), (\ref{eq:mr})
and discard the time evolution equation for $m$.

From Eq.~(\ref{eq:Phir}), we further notice that
the metric potential $\Phi$ is determined only up to
an additive constant. In GR, this freedom is commonly used to match
the outer edge of the computational domain to an external Schwarzschild
metric. This cannot be done in ST theories, as such theories do not
obey a direct analogue of the Birkhoff theorem
\cite{1923rmp..book.....B,1977JPhA...10..993K}. We therefore specify
a boundary condition for $\Phi$ using the method put forward by Novak \cite{1998PhRvD..57.4789N}:
$\Phi$ is constrained on the outer boundary of the computational domain
by requiring that
\begin{equation}
  K = \frac{e^{\Phi} }{\sqrt{1-\frac{2m}{r}}}
\label{KAN}
\end{equation}
is approximately constant in the weak-field regime,
far away from the star. 
$K$ is then evaluated for the initial profile (cf.~Sec.~\ref{initialprofiles}), fixed to be constant
during the evolution and determines $\Phi$ on the outer edge of the grid
$r=R_{\rm out}$ by inverting (\ref{KAN})
\begin{equation}
  \Phi(R_{\rm out})= \ln\left(K \sqrt{1-\frac{2m(R_{\rm out})}{R_{\rm out}}}
       \right)~.
  \label{KAN2}
\end{equation}
Note that the Birkhoff theorem in GR corresponds to the case $K=1$.
The error incurred from this procedure can be estimated by comparing
results obtained for different extents of the computational domain.
We obtain variations of order $|\Delta\varphi/\varphi|\sim 10^{-3}$
at the {radius of extraction} when the grid extent is decreased by a factor
2 (cf.\ Sec.~\ref{numerics} for more details on our numerical setup).
Similar errors are detected {in the collapse of a ST polytrope} if $K$ is set to 1, rather than evaluated
from the initial profile.

\subsubsection{Scalar-field equations.}
\label{sfeq}

The wave equation for the scalar field (\ref{jordanevolphi}) can
be written as a first-order system using the definitions (\ref{eq:psi})
and the identity $\partial_t \partial_r \eta = \partial_r \partial_t
\eta$ to obtain
\begin{align}
  \partial_t \varphi &= \alpha \psi~,
      \label{eq:varphit} \\
  \partial_t \eta &=\frac{1}{X} \partial_r \left( \alpha \psi \right) -  rX\alpha \eta (\eta \psi - 4\pi F\,S^r) + \frac{F_{,\varphi}}{2F} \alpha  \eta \psi~,
      \label{eq:etat2} \\
  \partial_t \psi &= \frac{1}{r^2X} \partial_r \left( \alpha r^2 \eta \right)
      +  rX\alpha \psi(\eta \psi - 4\pi F\,S^r)\notag\\
       &\quad- \frac{F_{,\varphi}}{2F} \alpha \psi^2 
      + 2\pi \alpha \left( {\tau} - {S}^r v + {D}
        - 3 \frac{P}{F^2} \right) F_{,\varphi}~.
      \label{eq:psit2}
\end{align}

In order to prescribe the behaviour of $\varphi$ at the outer
boundary, we consider the asymptotic behaviour of the scalar field at spatial infinity \cite{1992CQGra...9.2093D}%
\begin{align}
  \varphi(r)= \varphi_0 + \frac{\omega}{r} + \mathcal{O}\left(\frac{1}{r^2}
        \right)~,
  \label{varphiasym}
\end{align}
where $\varphi_0 = \mathrm{const}$ and
and $\omega$ denotes the \emph{scalar charge} of the star.
Physically,
we require that no radiation enters the spacetime from infinity
and therefore impose
an outgoing boundary condition \cite{1949pdep.book.....S} at spatial
infinity
\begin{equation}
  \lim_{r\to\infty}  \varphi(t,r) =  \varphi_{0}
        +\frac{f(t-r)}{r} + {\cal O}(r^{-2})\,,
\end{equation}
where $f$ is a free function of retarded time.
This condition can be translated into the following differential expressions
for $\eta$ and $\psi$,
\begin{align}
  &\partial_{t}\psi + \partial_{r}\psi + \frac{\psi}{r} = 0\,,
        \label{eq:OBCpsi} \\
  &\partial_{t}\eta + \partial_{r}\eta + \frac{\eta}{r}
     - \frac{\varphi-\varphi_{o}}{r^{2}} = 0\,,
        \label{eq:OBCeta}
\end{align}
and the scalar field $\varphi$ is directly obtained from
Eq.~(\ref{eq:varphit}). As shown in more detail
below [see Eq.~(\ref{Fab}) and the following discussion],
the value of $\varphi_0$
is degenerate with one of the parameters used to
describe the coupling function and is set to zero in our study
without loss of generality.

In practice, our computational domain extends to large but finite radii
and we approximate the physical boundary conditions by imposing
Eqs.~(\ref{eq:OBCpsi}), (\ref{eq:OBCeta}) at the outer edge rather
than at infinity. 
As already mentioned, we have tested the influence of the outer boundary location on our results and
observe only tiny variations of the order of $|\Delta\varphi/\varphi|\sim 10^{-3}$ in the extraction region when comparing with  simulations performed with  $R_{\rm out}$ twice as large.

\subsubsection{Matter equations in flux-conservative form.}
\label{mattereq}

The evolution equations (\ref{jordanevol}--\ref{jordanevolphi})
can be conveniently written in flux-conservative form
\cite{LeVeque1992,2008LRR....11....7F},
\begin{equation}
  \partial_t \textbf{U} + \frac{1}{r^{2}}\partial_{r} \left[ r^{2}
  \frac{\alpha}{X} \textbf{f(U)} \right] = \textbf{s(U)} \,, \label{vectorscl}
\end{equation}
where $\textbf{U}$ is the vector of the conserved variables
$\textbf{U}=[D,S^r,\tau]$ defined in Eqs.~(\ref{defD}-\ref{deftau}).
The fluxes $ \textbf{f(U)}=[f^{}_{{D}},f^{}_{{S}^r},f^{}_{{\tau}}]$ and the source $ \textbf{s(U)}=[s^{}_{{D}},s^{}_{{S}^r},s^{}_{{\tau}}] $ are
given by
\begin{align}
  f^{}_{{D}} &= {D} v\,, \label{eq:fD} \\
  f^{}_{{S}^r} &= {S}^r v + \frac{P}{F^2}\,, \\
  f^{}_{{\tau}} &= {S}^r-{D} v\,, \\
  s^{}_{{D}} &=-D\frac{F_{,\varphi}}{2F} \alpha ( \psi+\eta v)\,,
  \label{sourceD} \\
  s^{}_{{S}^r} &= ({S}^r v -{\tau} -{D}) \alpha X F
      \left(
        8\pi r \frac{P}{F^2} + \frac{m}{r^2}
        - \frac{F_{,\varphi}}{2F^2X}\eta
      \right)
      + \frac{\alpha X}{F} P \frac{m}{r^2}
      + 2\frac{\alpha P}{rX F^2} \nonumber \\
  &\quad -2r\alpha X {S}^r \eta \psi
     - \frac{3}{2}\alpha \frac{P}{F^2} \frac{F_{,\varphi}}{F} \eta
     - \frac{r}{2} \alpha X (\eta^2+\psi^2)
     \left( {\tau} + \frac{P}{F^2} + {D} \right) (1+v^2)\,, \\
  s^{}_{{\tau}} &= -\left( {\tau}+ \frac{P}{F^2} + {D} \right)
     r\alpha X
     \left[
       (1+v^2) \eta \psi + v (\eta^2 + \psi^2)
     \right]\,
     \notag
     \\ &\quad+\frac{\alpha}{2} \frac{F_{,\varphi}}{F}
     \left[
       {D} v \eta
       + \left( {S}^rv - {\tau}+ 3\frac{P}{F^2} \right) \psi
     \right]\,.
     \label{sourcetau}
\end{align}

Note that Ref.~\cite{2000ApJ...533..392N} misses a factor $1/a$ (in their notation) inside the argument of the radial derivative in their Eq.~(11). Inclusion of this factor and pulling the term proportional to $\eta v$ in Eq.~(\ref{sourceD}) out of the radial derivative enables us to cast the evolution equation for $D$ in the same form (\ref{vectorscl}) as the other matter equations. For the integration of the evolution equation for $D$, we therefore do not need the additional considerations  described in Sec.~2.1 of \cite{2000ApJ...533..392N}.

The hyperbolic structure of the system of equations (\ref{vectorscl}) is dictated by the Jacobian matrix of the fluxes \cite{2010nure.book.....B},
\begin{equation}
\textbf{J}^{}_{\textbf{U}} = \frac{\partial \textbf{f(U)}}{\partial \textbf{U}} \,.
\end{equation}
The characteristic speeds associated with the propagation of the matter fields are the eigenvalues $\lambda$ of $\textbf{J}^{}_{\textbf{U}}$,
\begin{align}
\lambda = & \left[  v, \frac{v+ c_{s}}{1+ v\,c_{s}},\frac{v- c_{s}}{1- v\,c_{s}} \right]\,.
\end{align}
Here {$c_{s} = \sqrt{(dP/d\rho)_S / h}$ ($S$ is the entropy)
is the local speed of sound given for our choice of EOS of the
form $P=P(\rho,\epsilon)$ by
\begin{equation}
h\,c^{2}_{s} = \frac{\partial P}{\partial \rho} + \frac{P}{\rho^{2}}\,\frac{\partial P}{\partial \epsilon}\,.
\end{equation}
The characteristic speeds are therefore exactly the same as in GR,
since they do not depend on the conformal factor
$F$. The high-resolution shock-capturing
scheme implemented in \textsc{gr1d}  for GR   \cite{2010CQGra..27k4103O}
can therefore be used in ST theories as well,
provided the conserved variables
$\textbf{U}$ and their fluxes $\textbf{f(U)}$ are generalised using
the expressions presented above.

\section{Physical setup}
\label{physicalsetup}
In this Section, we discuss in more detail the physical ingredients
entering our simulations. We discuss the EOS for the fluid
used in our work (Sec.~\ref{EOS}), the various choices for the
coupling function that relate the physical metric to its conformally
rescaled counterpart (Sec.~\ref{coupfun}) and the initial stellar
profiles used in our study (Sec.~\ref{indata}). We also provide
information on the quantities used to compare GW signals and detector
sensitivities in the context of monopole waves (Sec.~\ref{gwcurves}).

\subsection{Equation of state}
\label{EOS}
An EOS is required to close the hydrodynamical system of
equations. Specifically, it provides a prescription for the pressure
$P$ and other thermodynamic quantities as a function of the mass
density, internal energy (or temperature), and possibly the chemical
composition.  In this paper we study stellar collapse using the
so-called \emph{hybrid} EOS.  This EOS was introduced in
Ref.~\cite{1993A&A...268..360J} and qualitatively captures in closed
analytic form the 
expected stiffening of
the nuclear matter EOS
at nuclear density and
includes nonisentropic (thermal) effects to model the response of
shocked material. 
The hybrid EOS was
    widely used in early multi-dimensional core-collapse
simulations (e.g. \cite{1997A&A...320..209Z,2002A&A...393..523D}), and
the results of simulations using a hybrid EOS have been compared in
detail with those obtained with modern finite-temperature EOS; see
e.g.~Ref.~\cite{2007PhRvL..98y1101D,2008PhRvD..78f4056D}.

The hybrid EOS consists of a \emph{cold} and a \emph{thermal} part:
\begin{align}
  P= P_{\rm c} +P_{\rm th}\,.
\end{align}
The cold component $P_c$ is modelled in piecewise polytropic
form with adiabatic indices $\Gamma_1$ and $\Gamma_2$,
\begin{align}
  P_{\rm c} = 
  \begin{cases}
    K_1 \rho^{\Gamma_1} \qquad {\rm if}&\quad \rho\leq \rho_{\rm nuc}\,, \\
    K_2 \rho^{\Gamma_2} \qquad {\rm if}&\quad \rho> \rho_{\rm nuc}\,. \\
  \end{cases}
\end{align}
This expression
models both the pressure contribution from relativistic electrons,
which dominates at $\rho \le \rho_{\rm nuc}$,
and the stiffening at nuclear density due to the repulsive
character of the nuclear force.
The two components are matched at
``nuclear density'' which we set
to $\rho_{\rm nuc}= 2\times 10^{14} ~{\rm g/cm^3}$
following \cite{2002A&A...393..523D}.
We set $K_1=4.9345\times 10^{14}~{\rm [cgs]}$, as predicted for a
relativistic degenerate gas of electrons with electron fraction $Y_e=0.5$
\cite{1986bhwd.book.....S}, while $K_2=K_1 \rho_{\rm
nuc}^{\Gamma_1-\Gamma_2}$ is then obtained from requiring
continuity in $P$ at $\rho=\rho_{\rm
nuc}$. The specific internal energy follows from the first law of
thermodynamics applied to the case of adiabatic processes
\begin{align}
  \epsilon_{\rm c} = 
  \begin{cases}
  \frac{K_1}{\Gamma_1-1} \rho^{\Gamma_1-1} &\quad {\rm if}\quad
        \rho\leq \rho_{\rm nuc}\,, \\
  \frac{K_2}{\Gamma_2-1}  \rho^{\Gamma_2-1} + E_3   &\quad {\rm if}\quad
        \rho> \rho_{\rm nuc}\,,\\
  \end{cases}
\end{align}
where the integration constant $E_3$ is determined by continuity at
$\rho=\rho_{\rm nuc}$. The thermal contribution $P_{\rm th}$ is described by  a
$\Gamma$-law
with adiabatic index $\Gamma_{\rm th}$,
\begin{align}
  P_{\rm th} = (\Gamma_{\rm th} -1)\rho \epsilon_{\rm th}~,
\end{align}
where $\epsilon_{\rm th}=\epsilon - \epsilon_{\rm c}$ is the thermal
contribution to the internal energy, computed from the primitive
variable $\epsilon$. The flow is adiabatic  before bounce, implying
that $\epsilon\simeq\epsilon_{\rm c}$ and the total pressure is
described by considering only its cold contribution.
At core bounce, however,
the hydrodynamic shock results
in nonadiabatic flow and thus
triggers the onset of a non-negligible thermal contribution
to the EOS.

We consider a hybrid EOS characterised by three parameters:
$\Gamma_1$, $\Gamma_2$ and $\Gamma_{\rm th}$. The physical range of
these adiabatic indices has been explicitly studied in
Refs.~\cite{2007PhRvL..98y1101D,2008PhRvD..78f4056D}, where 2+1 GR
simulations of core collapse were used to compute the effective
adiabatic index of the finite-temperature EOS of Lattimer and Swesty
\cite{1991NuPhA.535..331L,1985NuPhA.432..646L} and Shen {\it et al.}
\cite{1998NuPhA.637..435S,1998PThPh.100.1013S}. 
In
  the collapse phase, electron capture decreases   
  the effective
adiabatic index below the value $\Gamma_1=4/3$ predicted for a
relativistic gas of electrons. More precisely,
comparisons
  with more detailed simulations yields a range from
$\Gamma_1\simeq
1.32$ to $\Gamma_1\simeq 1.28$
\cite{2007PhRvL..98y1101D,2008PhRvD..78f4056D,2011ApJS..197...20S}.
In particular, lower
values of $\Gamma_1$ are found when deleptonisation is taken into
account because electron capture onto nuclei before neutrino trapping
decreases $Y_{\rm e}$ for given $\rho$, thus softening the EOS.
Collapse is stopped by the stiffening of the EOS at nuclear density
which raises the effective adiabatic index $\Gamma_2$ above $4/3$.
Reference \cite{2008PhRvD..78f4056D} finds $\Gamma_2\simeq 3.0$ for
the Shen {\it et al.} EOS and $\Gamma_2\simeq 2.5$ for the
Lattimer-Swesty EOS. Finally, the thermal adiabatic index $\Gamma_{\rm
  th}$ models a mixture of relativistic and non-relativistic gas, and
is therefore physically bounded to $4/3<\Gamma_{\rm th}<5/3$.  We
select fiducial values $\Gamma_1=1.3$, $\Gamma_2=2.5$, $\Gamma_{\rm
  th}=1.35$ for our code tests presented in Sec.~\ref{numimpl}, and
explore a more extended parameter grid
around this model in
Sec.~\ref{results}.

\subsection{Coupling function}
\label{coupfun}

As introduced in Sec.~\ref{formulations}, ST theories with a single
scalar field and vanishing potential are described by a single free
function $F(\varphi)$. The phenomenology of ST theories is simplified,
however, by the fact that all modifications of gravity at
first PN order depend only on two parameters. These are the asymptotic values
of the first and second derivatives of $\ln F$
\cite{1992CQGra...9.2093D,1996PhRvD..54.1474D,1997PThPS.128..335C}\footnote{We
introduce factors $-1/2$ in Eq.~(\ref{defalpha0}),
and consequently a factor $-2$ 
in Eq.~(\ref{Fab}), to be consistent
with previous studies,
e.g.,~Refs.~\cite{1992CQGra...9.2093D,1993PhRvL..70.2220D,1996PhRvD..54.1474D}.},
\begin{align}
  \alpha_0 = -\frac{1}{2}\frac{\partial \ln F}{\partial\varphi}
        \bigg|_{\varphi=\varphi_0}\,,
\qquad  \beta_0 = -\frac{1}{2}\frac{\partial^2 \ln F}{\partial \varphi^2}
        \bigg|_{\varphi=\varphi_0}\,.
        \label{defalpha0}
\end{align}
The effective gravitational constant determining the attraction
between two bodies as measured in a Cavendish experiment is
\begin{align}
  \tilde G = G (1+\alpha_0^2)\,,
\end{align}
where $G$ is the bare gravitational constant entering the action.
Furthermore, the Eddington Parameterised post-Newtonian parameters
\cite{1923mtr..book.....E,1993tegp.book.....W} can be expressed exclusively
in terms of $\alpha_0$ and $\beta_0$ through
\begin{align}
  \beta^{\rm PPN} -1 = \frac{\alpha_0^2 \beta_0}{ 2 (1+\alpha_0^2)^2}\,,  \qquad
  \gamma^{\rm PPN} -1 = -2\frac{\alpha_0^2}{1+\alpha_0^2}\,. \label{eddppn}
\end{align}
For an interpretation of these
equations in terms of fundamental interactions, see
Ref.~\cite{1996PhRvD..53.5541D}.
In consequence, weak-field deviations from GR are completely determined by
the Taylor expansion of $\ln F$ to quadratic order
about $\lim_{r\to\infty}\varphi=\varphi_0$.
For these reasons, most of the literature on ST theories
has focused on coupling functions of quadratic form
\cite{1993PhRvL..70.2220D,1996PhRvD..54.1474D} and we follow this
approach by employing a coupling function
\begin{align}
  F= \exp \left[ -2  \alpha_0 (\varphi-\varphi_0)
        - \beta_0 (\varphi-\varphi_0)^2 \right]\,.
        \label{Fab}
\end{align}
The asymptotic value $\varphi_0$ does not represent an additional
degree of freedom in the theory
because it can be reabsorbed by a field redefinition $\varphi \to \varphi
+\varphi_0$
\cite{2007arXiv0704.0749D}
and we therefore set $\varphi_0=0$ without loss of
generality in the rest of this
paper.\footnote{The class of theories here parameterised by
$(\alpha_0,\beta_0)$ can equivalently be
represented using $F= \exp(-2 \beta_0 \varphi^2)$
but keeping $\varphi_0$ as an independent parameter, as done, e.g.,~in
Ref.~\cite{1993PhRvL..70.2220D}.} We can furthermore assume $\alpha_0\geq0$
because the sign of $\alpha_0$ is degenerate with the field
redefinition $\varphi \to -\varphi$. Despite its apparent simplicity,
this two-parameter family of ST theories is representative of all
ST theories with the same phenomenology up to first PN order. Brans-Dicke
theory  \cite{1961PhRv..124..925B} is a special case of Eq.~(\ref{Fab})
with the Brans-Dicke parameter [defined above Eq.~(\ref{eq:conformalTrafo})]
given by $\omega_{\rm BD}=
(1-6\alpha_0^2)/2\alpha_0^2$ and  $\beta_0=0$.
It is worth mentioning here that theories with the coupling function (\ref{Fab}) and strictly vanishing potential have been shown to exhibit non-viable cosmological evolutions \cite{1993PhRvD..48.3436D,1993PhRvL..70.2217D}; however, this can be cured by introducing a suitable (sufficiently flat) potential which leaves the phenomenology on stellar scales unchanged \cite{2008PhRvD..78h3530J,2014PhRvD..90l4091S}.

It is well known that
all deviations in the structure
of spherically symmetric bodies in ST theory from their general relativistic
counterparts are given in terms of a series of
PN terms proportional to $\alpha_0^2$ \cite{1992CQGra...9.2093D,2004AIPC..736...35E};
cf.~also Eq.~(\ref{eddppn}) above.
Any ST theory with $\alpha_0=0$ is therefore
perturbatively equivalent to GR and current observations (see below)
constrain $\alpha_0$ to be very small.
In 1993, however, Damour and Esposito-Far\`ese
\cite{1993PhRvL..70.2220D,1996PhRvD..54.1474D} discovered a remarkable
non-perturbative effect called \emph{spontaneous scalarisation},
which introduces macroscopic modifications to the structure of
NSs  even
when $\alpha_0$ is very small or vanishes \cite{2011JCAP...08..027H}.
For certain values $\beta_0<0$, there exists a threshold in the
compactness ($M/R$, where $M$ is the total mass of the object and $R$ is its radius)  of stellar structure at which
spherically symmetric equilibrium solutions develop significant scalar
hair. One can find three distinct solutions in this regime: besides a
weakly scalarised
solution where the ratio between the scalar charge and the star's
mass $\omega/M$ is of the order of $\alpha_0$, two strong-field
solutions appear where this ratio is of order unity
\cite{1998PhRvD..58f4019N,1998PhRvD..57.4802H}. If $\alpha_0=0$,
the weak-field solution is a GR star and the two strong field
solutions coincide. Notably, scalarised solutions are present for
compactness values of order $M/R\gtrsim0.2$ \cite{2004AIPC..736...35E}, as realised in NSs.
When present, scalarised neutron stars can be energetically
favoured over their weak-field counterparts
\cite{1993PhRvL..70.2220D,1996PhRvD..54.1474D,1997PThPh..98..359H}, allowing for the
possibility of dynamical transitions between the two branches of
solutions \cite{1998PhRvD..58f4019N}. Spontaneously scalarised stars
have been found for $\beta_0\lesssim -4.35$
\cite{1998PhRvD..58f4019N,1998PhRvD..57.4802H}, but the exact value
of this threshold depends on the EOS.

\begin{figure}[t]\centering
\includegraphics[width=0.95\textwidth]{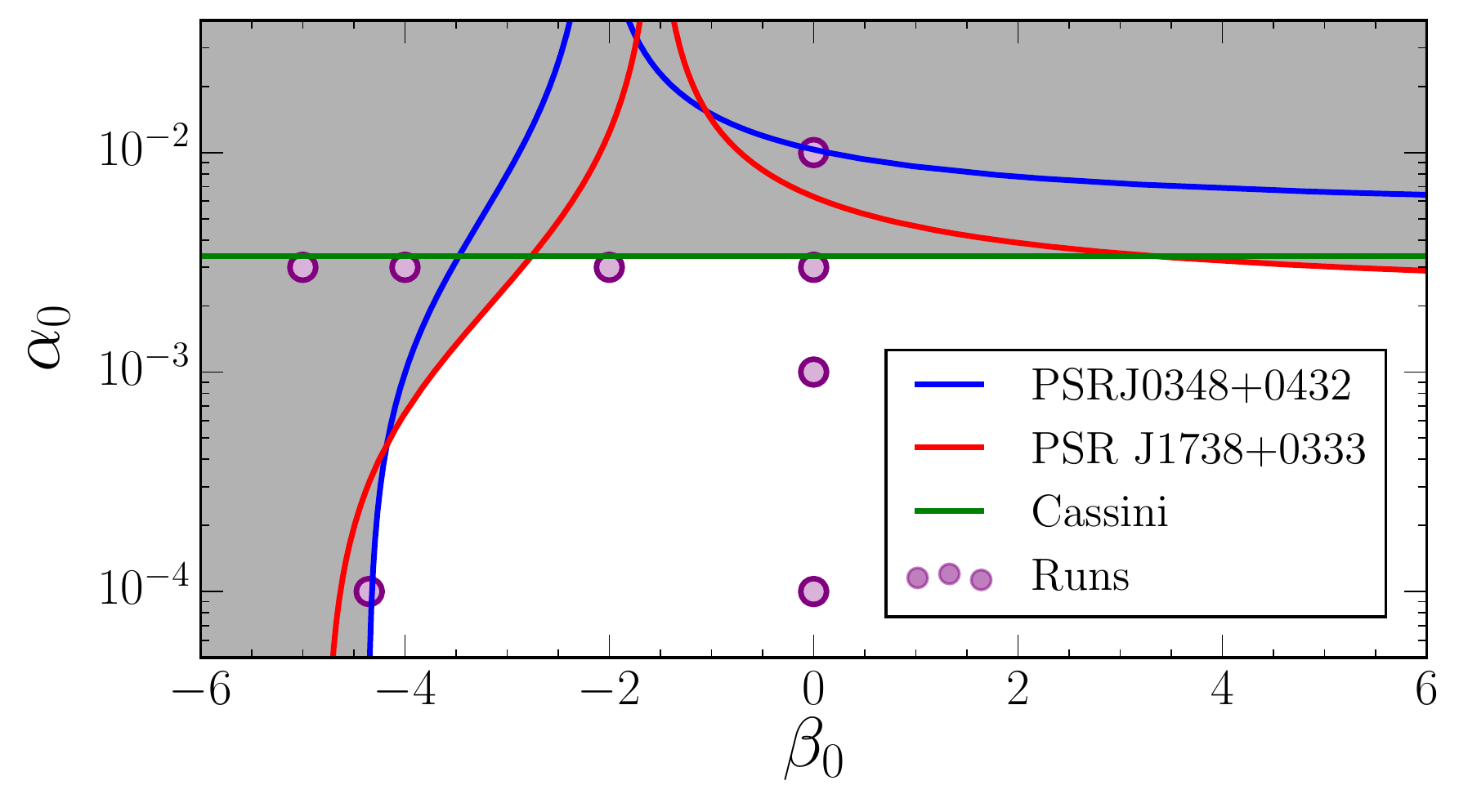}
\caption{Experimental constraints on the ST-theory parameters
$(\alpha_0,\beta_0)$ entering the coupling function $F$. The shaded
area is currently ruled out by observations; GR lies at
$\alpha_0=\beta_0=0$. The most stringent constraints on $\alpha_0$
are provided by the Cassini space mission while the binary-pulsar
experiments impose strong bounds on $\beta_0$. Circles mark our choices
for $(\alpha_0,\beta_0)$ used in Sec.~\ref{results}.
This figure is produced using the data published in Fig.~6.3
of Ref.~\cite{2015CQGra..32x3001B}. }
\label{obsconstraints}
\end{figure}
The $(\alpha_0,\beta_0)$ parameter space of ST theories has been
severely constrained by observations. Solar System probes include
measurements of Mercury's perihelion shift \cite{1990grg..conf..313S},
Lunar Laser Ranging \cite{2009IJMPD..18.1129W}, light deflection
measured with Very-Long-Baseline Interferometry \cite{2004PhRvL..92l1101S},
and the impressive bound $\alpha_0<3.4 \times 10^{-3}$  obtained
with the Cassini space mission \cite{2003Natur.425..374B}. Timing
of binary pulsars currently provides the tightest constraints in
the $\beta_0$ direction of the parameter space \cite{2014arXiv1402.5594W}.
In particular, observations from pulsars PSR J1738+0333
\cite{2012MNRAS.423.3328F} and PSR J0348+0432 \cite{2013Sci...340..448A}
(both orbiting a white dwarf companion) rule out a wide range of theories
exhibiting prominent spontaneous scalarisation. Current observational
constraints are summarised in Fig.~\ref{obsconstraints} where the
shaded area is now excluded. Note, however, that the binary-pulsar
constraints apply
to the case of a single massless scalar field. Scalar-tensor theories
with multiple scalar fields \cite{2015CQGra..32t4001H} or
with one massive field \cite{2016PhRvD..93f4005R} may still
lead to spontaneously scalarised neutron stars over a wide range
of the theories' parameters without coming into conflict with the
binary pulsar observations.

\subsection{Initial profiles}
\label{indata}

We perform simulations of stellar collapse starting from
two types of initial data: (i) polytropic models generated in the
static limit of the ST theory equations and (ii) ``realistic''
SN progenitors obtained from stellar evolutionary
computations performed by Woosley \& Heger \cite{2007PhR...442..269W}.
\begin{itemize}
\item[(i)] In the static limit, the evolution equations presented
           in Sec.~\ref{metriceq}--\ref{mattereq} reduce to
           (cf.~\cite{1993PhRvL..70.2220D,2015CQGra..32t4001H})
\begin{align}
  \partial_r \Phi &= FX^2 \left(
      \frac{m}{r^2}+4\pi r \frac{P}{F^2} + \frac{r}{2F}\eta^2
      \right)\,, \label{eq:staticPhir} \\
  \partial_r m &=
      4\pi r^2 \frac{\rho h-P}{F^2} + \frac{r^2}{2F}\eta^2\,, \\
  \partial_r P &=
      -\rho h FX^2 \left(
      \frac{m}{r^2} + 4\pi r \frac{P}{F^2} + \frac{r}{2F} \eta^2
      \right)
      + \rho h \frac{F_{,\varphi}}{2F}X\eta\,, \\
  \partial_r \varphi &= X \eta\,, \\
        \partial_r \eta &=
      -2\frac{\eta}{r} - 2\pi X \frac{\rho h - 4P}{F^2} F_{,\varphi}
      -\eta FX^2 \left(
      \frac{m}{r^2} + 4\pi r \frac{P}{F^2} + \frac{r}{2F} \eta^2
      \right)
      + \frac{X}{2} \frac{F_{,\varphi}}{F} \eta^2\,,
      \label{eq:staticetar}
\end{align}
which generalise the Tolman–Oppenheimer–Volkoff
\cite{1939PhRv...55..364T,1939PhRv...55..374O} equations to ST
theory. As in GR, the equation for the metric potential $\Phi$
decouples from the remainder and we need an EOS
$P=P_{\rm EOS}(\rho)$ to close the system.

In practice, we integrate the system
(\ref{eq:staticPhir})--(\ref{eq:staticetar}) outwards starting at the origin
where boundary and regularity conditions require
\begin{align}
  \Phi(0)\;=\;0\, , 
  \quad
  m(0)\;=\;0\,,
  \quad
  P(0)\;=\; P_{\rm EOS}(\rho_c)\,,
  \quad
  \varphi(0)\;=\; \varphi_c \,,
  \quad
  \eta(0)\;=\; 0 \,.
  \label{eq:staticBCs}
\end{align}
Here, $P_c$ (or, alternatively, $\rho_c$) is a free parameter determining
the overall mass and size of the star and the central value of the
scalar field $\varphi_c$ is related through the integration to the
value of $\varphi$ at infinity. In our case, the boundary condition for the
scalar field is $\varphi(r\rightarrow \infty)=\varphi_0=0$ and the task is
to identify the ``correct'' central value $\varphi_c$ that satisfies the
outer boundary condition. From a numerical point of view, this task
represents a {\em two point boundary value problem} \cite{1989nrca.book.....P}
and we use a {\em shooting algorithm} to solve it.
For this purpose, we note that
the integration terminates at the stellar surface
$r_s$ defined as the innermost radius where $P=0$.
From this radius $r_s$, we could in principle continue the integration
to infinity by setting the matter sources to zero and switching to a
compactified radial coordinate such as $y\equiv 1/r$. We have found
such a scheme to work successfully \cite{2015CQGra..32t4001H},
but here we implement an equivalent, but conceptually simpler
algorithm. The numerical solution computed for $r\leq r_s$
can be matched to a vacuum solution at $r>r_s$ to relate the scalar
field at the stellar
surface $\varphi_s$ to its asymptotic value $\varphi_0$
at $r=\infty$ \cite{1993PhRvL..70.2220D}:
\begin{equation}
  \varphi_s = \varphi_0 -
      \frac{X_s \eta_s}{\sqrt{(\partial_r \Phi_s)^2 + X_s^2 \eta_s^2}}
      \mathrm{arctanh}
      \frac{\sqrt{(\partial_r \Phi_s)^2 + X_s^2 \eta_s^2}}
      {\partial_r \Phi_s + 1/r_s}\,, \label{eq:phi0B}
\end{equation}
where the subscript $s$ denotes quantities evaluated at $r_s$.
The shooting algorithm starts the integration of
Eqs.~(\ref{eq:staticPhir}--\ref{eq:staticetar}) with
some initial guess $\varphi(0)$, obtains the corresponding
$\varphi_s$ and then iteratively improves the choice
of $\varphi(0)$ until it leads to a $\varphi_s$ that
satisfies Eq.~(\ref{eq:phi0B}) within some numerical tolerance
($10^{-10}$ for the absolute difference in our case). 

The central density or pressure can be freely chosen
and parameterises the family of static solutions
for a given ST theory $(\alpha_0,\beta_0)$ in the same way as it does in GR.
The members of this one-parameter family of solutions are often
characterised by their
total gravitational mass
which is given by \cite{1993PhRvL..70.2220D}
\begin{equation}
  m_{\rm \infty} = r_s^2 \partial_r \Phi_s \sqrt{1-\frac{2m_s}{r_s}}
      \exp \left[
        -\frac{\partial_r \Phi_s}{\sqrt{(\partial_r \Phi_s)^2 + X_s^2\eta_s^2}}
      \mathrm{arctanh}
        \frac{\sqrt{(\partial_r \Phi_s)^2+X_s^2 \eta_s^2}}{\partial_r \Phi_s
        + 1/r_s}
      \right].
  \label{eq:Mgrav}
\end{equation}
All polytropic initial profiles used in this work
are generated using a polytropic EOS $P=K
\rho^\Gamma$ with $K=4.9345\times 10^{14}~{\rm [cgs]}$, $\Gamma=4/3$
and central mass density $\rho_c=10^{10}\,{\rm g~cm^{-3}}$; these
parameters are considered qualitatively reasonable approximations
to model iron cores supported by the degeneracy
pressure of relativistic electrons
\cite{1986bhwd.book.....S}. In particular, the choice $\rho_c=10^{10}\,{\rm g~cm^{-3}}$ results in stars with baryonic mass $\sim 1.44 M_{\odot}$, slightly below the Chandrasekhar limit \cite{1990RvMP...62..801B}.

\item[(ii)]
We also perform core collapse simulations using more realistic pre-SN
models. Woosley and Heger \cite{2007PhR...442..269W} evolved
non-rotating single stars up to the point of
iron core collapse
\cite{1978ApJ...225.1021W,2002RvMP...74.1015W}.  Here, we consider two
specific models of their catalogue obtained from the evolution of
stars with zero-age-main-sequence (ZAMS) mass $M_{\rm ZAMS}=12
M_\odot$ and $40 M_\odot$. We refer to these models as WH12 and WH40
respectively. 
 Model WH12 has a steep density gradient
  outside its iron core, which results in a low accretion rate after
  bounce. Even if no explosion occurs, the delay to BH formation would
  be multiple seconds and no BH forms over the time we simulate. Model
  WH40, on the other hand, has a very shallow density gradient,
  resulting in a high accretion rate after bounce. This pushes the
  protoneutron star over its maximum mass and leads to BH formation
  within a few hundred milliseconds of bounce
(cf.~\cite{2011ApJ...730...70O}). Hence, we use model
  WH12 to explore ST theory for a scenario in which core collapse
  results in a stable NS and model WH40 for a scenario in which the
  protoneutron star collapses to a BH.
   
Since WH12 and WH40 are Newtonian
models, we initialise the scalar-field variables $\varphi$, $\psi$ and
$\eta$ to $0$. An unfortunate consequence of this artificial (but
unavoidable) approximation is that no scalar-field dynamics occur at
all if $\alpha_0=0$:
 all source terms on the right-hand side of
Eqs.~(\ref{eq:varphit}--\ref{eq:psit2}) vanish at all times, and the
evolution proceeds exactly as in GR. We overcome this problem, by
using small but non-zero values for $\alpha_0$, which triggers a brief
initial transient
in the scalar field that
afterwards settles down into a smooth but non-trivial
configuration eventually leading to significant scalar field dynamics
as the collapse progresses through increasingly compact stages of the core.
\end{itemize}
\begin{figure}
\includegraphics[clip=true,width=\textwidth]{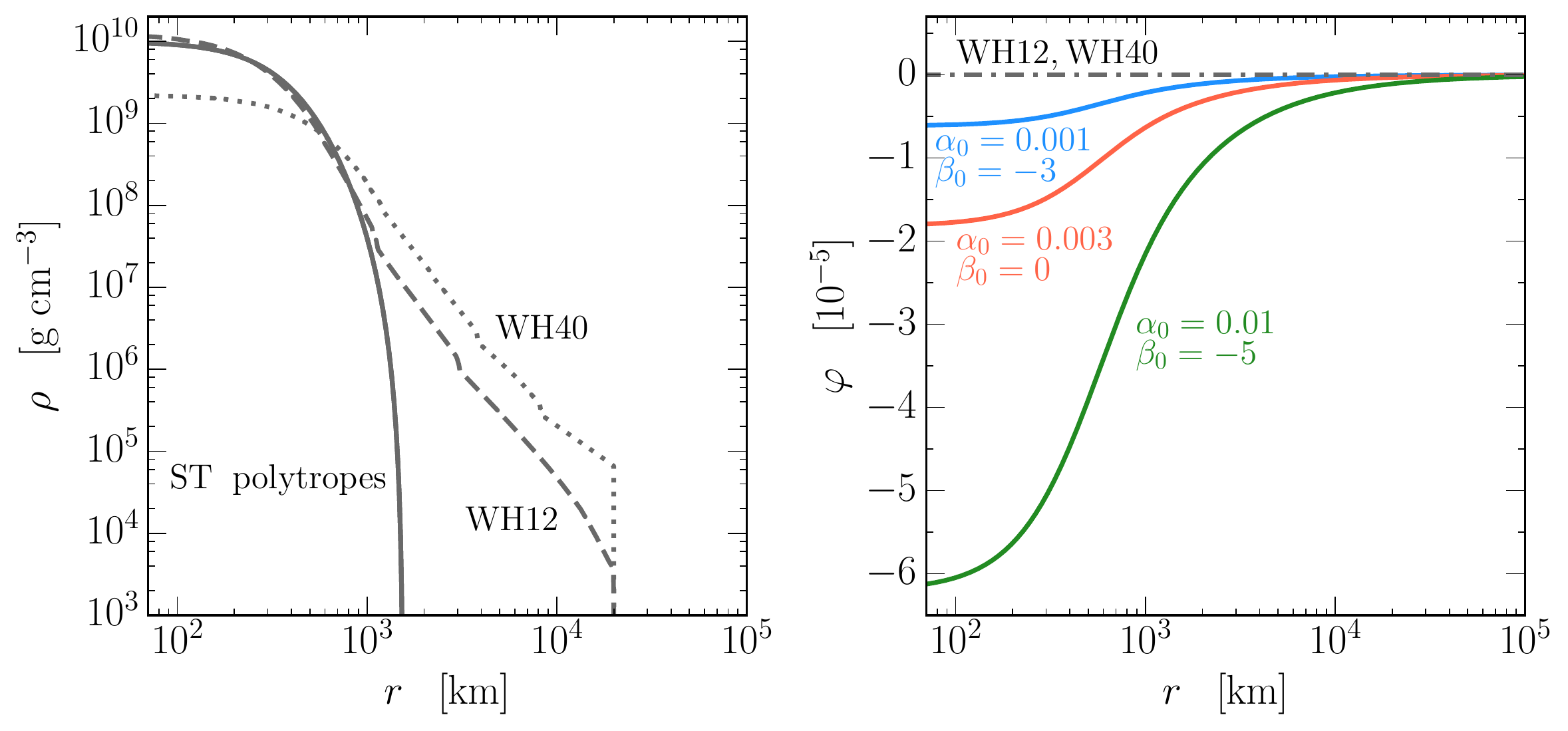}
\caption{ Mass-density (left panel) and scalar-field (right panel)
  profiles for the initial data considered in this study. In
  particular, dashed and dotted lines show the $M_{\rm ZAMS}=12
  M_\odot$ (WH12) and $M_{\rm ZAMS}=40 M_\odot$ (WH40) pre-SN models
  of Woosley and Heger \cite{2007PhR...442..269W} while the solid
  lines show three $\Gamma=4/3$ polytropes generated in ST theories
  with $(\alpha_0,\beta_0)=(0.001,-3),(0.003,0),(0.01,-5)$. The
  mass-density distributions of all three ST polytropes are
  indistinguishable from their GR counterparts.  The more realistic
  models WH12 and WH40 mostly differ from the polytropic ones through
  the presence of outer low-density layers.
    Note that
    we cut the WH models at $r_\mathrm{s} = 2\times 10^4\,\mathrm{km}$
    and pad them with an artificial atmosphere of $\rho_\mathrm{atm} =
    1\,\mathrm{g}\,\mathrm{cm}^{-3}$.
     The scalar field is more
  pronounced in models with higher $\alpha_0$, but the low compactness
  of these models prevents spontaneous scalarisation.
  The scalar field is initialised
  to zero when the WH models are evolved.}
\label{initialprofiles}
\end{figure}
For both classes of initial data there remains one degree of freedom that
we need to specify: the metric function $\Phi$ is determined by
Eq.~(\ref{eq:staticPhir}) only up to an additive constant. While our
integration in case (i) starts with $\Phi(0)=0$, we can trivially shift the profile
of $\Phi(r)$ by a constant (leaving all other variables unchanged) and
still have a solution 
of the system of
Eqs.~(\ref{eq:staticPhir})-(\ref{eq:staticetar}).
We use this freedom to
enforce that the physical metric component $g_{tt}=1$ as
$r\rightarrow \infty$, so that coordinate time is identical to
the proper time measured by an observer at infinity. In practice,
this is achieved by using very large grids and fitting
$\Phi = \Phi_0 + \Phi_1/r$ on the outer parts. $\Phi_0$ is then
the constant we subtract from the entire profile $\Phi(r)$.
The realistic initial models of case (ii) above are calculated without
a scalar field and in that case our procedure is equivalent to the
standard matching in GR based on the Birkhoff theorem.

For illustration, we show in Fig.~\ref{initialprofiles}
some of the initial profiles used in this study.
Because of the low
compactness of iron cores, the polytropic profiles
for all values of $\alpha_0 \le 0.01$
present very
similar mass-density distributions which also very closely resemble
their GR counterpart.  The magnitude of the scalar field inside the
star increases as larger values are chosen for $\alpha_0$ (cf.~right
panel of Fig.~\ref{initialprofiles}) while outside the star $\varphi$
rapidly approaches the $1/r$ behaviour of Eq.~(\ref{varphiasym}).  In
the left panel of Fig.~\ref{initialprofiles}, we also see that the
realistic pre-SN models WH12 and WH40 are well approximated by a
$\Gamma=4/3$ polytrope
in their central regions
$r\lesssim10^3~{\rm km}$; outer 
less degenerate
layers of lighter elements, however, substantially
broaden the mass-density distribution outside the iron core.

In order to overcome instabilities arising in our numerical scheme
due to zero densities $\rho$ \cite{2010CQGra..27k4103O},
we add an artificial atmosphere outside the stellar surface
$r_s$.
More specifically, we pad the polytropic profiles with a layer of
constant mass density $\rho_{\rm atm}=1~{\rm g~cm}^{-3}$. The WH
models are cut at
$r_{\rm s}=2\times 10^4~{\rm km}$ (cf.~Fig.~\ref{initialprofiles})
and padded with an artificial atmosphere of $\rho_{\rm atm}=1~{\rm
g~cm}^{-3}$. By comparing evolutions using different values for the
atmospheric density, we find the atmosphere to be completely irrelevant
to the dynamics of the star, but we observe that
significantly larger values than
$\rho_{\rm atm}=1~{\rm g~cm}^{-3}$ unphysically affect
the propagation of the scalar wave signal such that the wave signal
does not converge in the limit of large extraction radii.
We estimate the resulting error for our choice by comparison
with otherwise identical simulations using
instead $\rho_{\rm atm}=10~{\rm g~cm}^{-3}$;
the observed differences are $|\Delta h(t)|/h(t)\sim 0.3\%$ in the
extracted waveform [cf. Eq.~(\ref{monopoleht}) below]. 

\subsection{GW extraction and detector sensitivity curves}
\label{gwcurves}
The output of a GW detector $s(t)=n(t)+h(t)$ is the sum of  noise
$n(t)$ and  signal $h(t)$. For quadrupole GWs, as present in GR,
$h(t)$ is related to the metric perturbation $h_{+,\times}$ in the
transverse traceless gauge  through the beam pattern
functions
$A_{+,\times}$: $h(t)= A_+ h_+(t) +  A_\times h_\times(t)$
\cite{1973grav.book.....M}.
Monopole GWs are present in  ST theory
and are related to the dynamics of the scalar field $\varphi$. In this case, the detector response $h(t)= A_\circ h_\circ (t)$ is given by the metric perturbation $h_\circ (t)$  weighted by the correspondent beam pattern $A_\circ$ \cite{2009PhRvD..79h2002N,2013LRR....16....9Y}\footnote{The most sensitive directions corresponding to $A_+$, $A_\times$ and $A_\circ$ are all different from each other. If only these three polarisations are present, a network of four detectors can in principle disentangle them and estimate the source direction \cite{1994PhRvD..50.7304S}. Note also that optimally oriented sources correspond to $A_+=1$, $A_\times=1$ but $A_\circ=1/2$. \cite{2009PhRvD..79h2002N,2013LRR....16....9Y}.}.
 If we denote by $\tilde h(f)$  and $\tilde n(f)$ the
Fourier transform of $h(t)$ and $n(t)$, respectively, the (one-sided)
noise power spectral density  $S_n(f)$ is defined as
\begin{align}
  \langle \tilde n(f) \tilde n^*(f') \rangle = \frac{1}{2} \delta(f-f') S_n(f)\,,
\end{align}
where $\langle\cdot\rangle$ denotes a time average for stationary
stochastic noise. The signal-to-noise ratio is defined as (see
\cite{2015CQGra..32a5014M} where the numerical factor is derived;
see also \cite{2009LRR....12....2S})
\begin{align}
  \rho^2 =\int_0^\infty \frac{4 |\tilde h(f)|^2}{S_n(f)} df\,.
\end{align}
The characteristic strains for noise and signal are defined as
\begin{align}
  h_n(f)=\sqrt{f\,S_n(f)}\,, \qquad 
  h_c(f)= 2 f |\tilde h(f)|
\end{align}
such that $\rho^2$ can be written as the squared ratio
between signal and noise:
\begin{align}
  \rho^2 =\int_{-\infty}^{+\infty} \left[ \frac{h_c(f)}{h_n(f)}
        \right]^2 d\ln f\,.
\end{align}
The most common convention used to display detector sensitivity curves
involves plotting the square root of the power-spectral density
\begin{align}
  \sqrt{S_n(f)}= \frac{h_n(f)}{\sqrt{f}}\,;
\end{align}
and the analogous quantity \cite{2015CQGra..32a5014M}
\begin{align}
  \sqrt{S_h(f)}= \frac{h_c(f)}{\sqrt{f}} = 2 \sqrt{f}|\tilde h(f)|\,,
  \label{sourcestrain}
\end{align}
which characterises the GW signal.\footnote{The convention for $\sqrt{S_h(f)}$ used in Ref.~\cite{1998PhRvD..57.4789N} differs by a factor $2$ when compared to those of Ref.~\cite{2015CQGra..32a5014M} used here.}
In the following we will use sensitivity curves $\sqrt{S_n(f)}$ for:
\begin{itemize}
  \item[(i)] the Advanced LIGO detectors  \cite{2015CQGra..32g4001T,2016arXiv160203838T}  in their zero-detuned
  high-power configuration, as anticipated in
  \cite{LIGO-aLIGODesign-Sensitivity};
  \item[(ii)] the proposed  Einstein Telescope \cite{2010CQGra..27s4002P}, using the analytic
  fit reported in \cite{2009LRR....12....2S}.
\end{itemize}
Scalar waves are also promising sources for future GW experiments targeting the deci-Hertz regime, such as the proposed space mission DECIGO \cite{2011CQGra..28i4011K}. 

In contrast to GR, ST theories admit gravitational radiation
in spherical symmetry, specifically in the form of
a radiative monopole of the scalar field or,
equivalently, a so-called {\em breathing mode} when
considering the 
Jordan frame. 
The metric perturbation  of a monopole scalar wave in ST theory is \cite{1992CQGra...9.2093D}
\begin{align}
 h_\circ(t)=\frac{2}{D} \alpha_0 r(\varphi-\varphi_0)\,,
  \label{monopoleht}
\end{align}
where $D$ is the distance between the detector and the source and
the scalar field $\varphi$ is evaluated at radius $r$. The
factor $\alpha_0$ is due to the coupling between the scalar field
and the detector and limits
the potential of GW observations to constrain ST theories
\cite{2014PhRvD..89j4059B}. Throughout this paper we consider optimally oriented sources, such that $h(t)= A_\circ h_\circ(t)= h_\circ(t) /2 $ \cite{2009PhRvD..79h2002N,2013LRR....16....9Y}.

In analysing our simulations, we proceed as follows.
At a given radius $r_{\rm ext}$,
we extract $\varphi(t)$ and compute $h(t)$. 
In order to eliminate the brief unphysical transient (cf.~Sec.~\ref{numerics}),
we truncate this early part from the time domain waveform $h(t)$.
We then obtain $\tilde h(f)$
numerically using a Fast Fourier Transform (FFT)
algorithm.
To reduce spectral leakage, the FFT algorithm is applied to data $h(t)$ mirrored about the latest timestep available and the result $\tilde h(f)$ is normalised accordingly \cite{2013sdmm.book.....I}. This confines spectral leakage to frequencies $f \gtrsim 200$ Hz (cf.\ the tails in Fig.~\ref{spanSTa0} and \ref{spanSTb0}) where the signal is very weak. Finally, we compute $\sqrt{S_h(f)}$ from
Eq.(\ref{sourcestrain}) and compare it with the detectors' sensitivity curves $\sqrt{S_n(f)}$.

\section{Numerical implementation}
\label{numimpl}
In this Section, we provide details of our numerical scheme, stressing the modifications needed in ST theories with respect to the GR version of the code  (Sec~\ref{numerics}). We present the convergence properties in Sec.~\ref{convergence}. 

\subsection{Second-order finite differences and high-resolution shock capturing}
\label{numerics}

Our numerical code is built on top of {\sc gr1d}, an open-source
spherically-symmetric {\sc Fortran} 90+ code developed by O'Connor and
Ott \cite{2010CQGra..27k4103O}.  {\sc gr1d} has been applied to a
range of problems in stellar collapse and BH formation (e.g.,
\cite{2011ApJ...730...70O,2013ApJ...762..126O}). Its most recent GR version is available at
\cite{OTTwebsite} and includes energy-dependent
neutrino radiation transport \cite{2015ApJS..219...24O}.

As in the GR case, the constraint equations (\ref{eq:Phir}) and
(\ref{eq:mr}) for the metric functions $\Phi$ and $m$ are integrated
using standard second-order quadrature. In the scalar-field equations
(\ref{eq:varphit}--\ref{eq:psit2}), the source terms are discretised
using centred second-order stencils.  Due to the potential formation
of shocks in the matter variables, their evolution is handled with
a high-resolution shock capturing scheme as described in detail in
Sec.~2.1 of \cite{2010CQGra..27k4103O}. For our evolutions in ST
theory, we extended the flux and source terms of {\sc gr1d} in
accordance with our Eqs.~(\ref{eq:fD})--(\ref{sourcetau}). Integration
in time of the evolution equations for the matter and scalar fields is
performed using the method of lines with a second-order Runge-Kutta
algorithm. One significant difference from the GR case arises from the
presence of the scalar field as a dynamical degree of freedom with the
characteristic speed of light, whereas in spherical symmetry
in GR we only have to consider
the characteristic speed of sound for the matter degrees of
freedom (cf. Sec.~\ref{mattereq}).
In order to satisfy the Courant-Friedrichs-Lewy
stability condition we therefore determine the timestep using
the speed of light instead of the speed of sound, which results in
smaller values for the allowed timestep as compared with the corresponding
evolutions in GR.

As discussed in Sec.~\ref{mattereq}, a key ingredient in the
implementation of shock capturing methods is the use of a system of
evolution equations in flux conservative form which is available in
terms of the conserved variables $D,S^r,\tau$ but not in the primitive
variables $\rho$, $v$ and $\epsilon$. The primitive variables appear
in the constraint equations for the metric, the flux terms of the
shock-capturing scheme, in the EOS, and also form convenient
diagnostic output.  Conversion between the two sets of variables is
thus required at each timestep.  This process is straightforward for
the direction primitive $\rightarrow$ conserved;
cf.\ Eqs.~(\ref{defD}--\ref{deftau}).  The reverse transformation,
however, is non-trivial because of the presence of the pressure $P$
which is an intrinsic function of $\rho$ and $\epsilon$ given by the
EOS. This conversion is performed iteratively using a Newton-Raphson
algorithm: given an initial guess $\hat P$ for the pressure from the
previous timestep, we first calculate in this order
\begin{align}
  v &= \frac{{S}^r}{{\tau} + \hat P/F^2 + {D}}\,,
      \label{eq:vofP} \\
  \rho &= \frac{F^{3/2} {D}}{X}\sqrt{1-v^2}\,, \label{eq:rhoofP} \\
  \epsilon &= h - \frac{\hat P}{\rho} -1
      = \frac{F^2({\tau} + {D})(1-v^2) - \hat P v^2}{\rho } - 1\,.
      \label{eq:epsofP}
\end{align}
Then we compute an updated estimate for the pressure from the EOS
$P=P(\rho,\epsilon)$, and iterate this procedure until convergence.

The evolution of the scalar field turns out to be susceptible to
numerical noise near the origin $r=0$, which represents
a coordinate singularity. In order to obtain long-term stable
evolutions, we add artificial dissipation terms of Berger-Oliger type
\cite{1995tpdm.book.....G} to the scalar evolution equations.
Specifically, we add a dissipation term of the form
$\mathcal{D}\times \Delta r^4 \times \partial^4 u /\partial r^4$
to the right-hand-side of Eqs.~(\ref{eq:varphit})--(\ref{eq:psit2}),
where $u$ stands for either of the scalar-field variables, $\Delta
r$ is the width of the grid cell, and $\mathcal{D}$ is a dissipation
coefficient. In practice, we obtain good results using
$\mathcal{D}=2$. 

In all our simulations, the grid functions exhibit much stronger
spatial variation in the central region of the star than in the
wave zone. In order to accommodate these space dependent requirements
on the resolution of our computational domain,
we use a numerical domain composed of an inner grid with constant  and an
outer grid with logarithmic spacing. This setup enables us to capture the
dynamics of the inner core with high accuracy while maintaining
a large grid for GW extraction at tolerable computational
cost. Unless specified otherwise, we use the following grid setup.
The outer edge of the grid is placed at $R_{\rm out}=1.8
\times 10^5 {~\rm km}$ and the two grid components are matched at
$R_{\rm match}=40 {~\rm km}$. The cell width of the inner grid is
$\Delta r=0.25 {~\rm km}$. The total number of zones is set to
$N_{\rm zones}=5000$, such that 160 (4840) zones are present in the
inner (outer) grid. 4 ghost cells are added at both $r=0$ and
$r=R_{\rm out}$ for implementing symmetry and boundary conditions.
GW signals are extracted at $r_{\rm ext}=3\times
10^{4}~{\rm km}$, which is well outside the surface of the star but
sufficiently far from the outer edge of the grid $R_{\rm out}$ to
avoid contamination from numerical noise 
from
the outer boundary. We simulate the evolution for $0.7$ s to allow for the entire GW signal to cross the extraction region.

Radial gauge, polar slicing coordinates are not well adapted to BH
spacetimes: as the star approaches BH formation, the lapse function
$\alpha$ tends to zero in the inner region \cite{2008itnr.book.....A}
and inevitably introduces
  significant
numerical noise. The stellar evolution, however, is
effectively frozen as $\alpha\to 0$. Following Novak
\cite{1998PhRvD..57.4789N}, we handle BH formation by explicitly
stopping the evolution of the matter variables while we let the scalar
field propagate outwards. In practice, we freeze the matter evolution
whenever the central value of $\alpha$ becomes smaller than
$\alpha_T=5\times 10^{-3}$.  By varying the threshold $\alpha_T$ over
two orders of magnitude, we verified this procedure introduces a
negligible error $|\Delta\varphi|/\varphi\lesssim 1\%$ on the
extracted wave signal in case of BH formation.

A final note on the numerical methods concerns the time window used
for the wave extraction. As mentioned in Sec.~\ref{initialprofiles},
our initial data for the realistic progenitor 
models
require us to trigger scalar dynamics by using a small
but non-zero value for $\alpha_0$ 
that induces a brief transient in the wave signal. This
transient is not part of the physical signal we are interested in and
is removed by calculating waveforms in an interval starting not at
zero retarded time, but shortly afterwards: we use for this purpose
the time window $[t_i,t_f]$ with \mbox{$t_{i}=r_{\rm ext}/c+ 0.006$ s}
to \mbox{$t_{f}=r_{\rm ext}/c + 0.6$
s} from the beginning of the simulation. 
This provides us with waveforms of
total length $\Delta t\sim 0.6$ s
corresponding to a lower bound  $f\sim 1.7 $ Hz  in the frequency
domain. Note that our waveforms are significantly longer than
those obtained in previous studies of collapse in ST theories
\cite{2000ApJ...533..392N}.

\subsection{Self-convergence test}
\label{convergence}
\begin{figure}
\includegraphics[width=\textwidth]{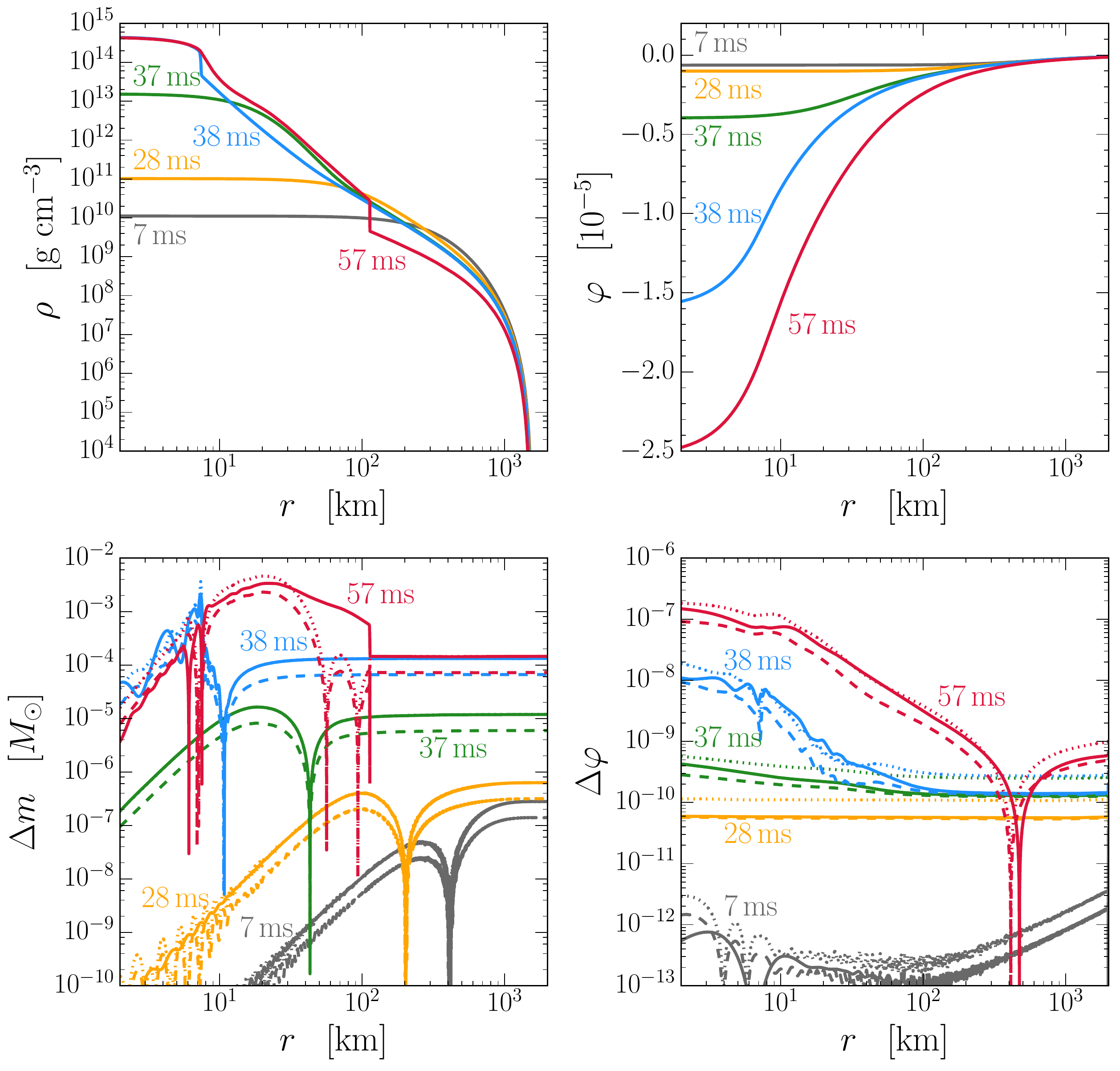}
\caption{Convergence test of stellar collapse in ST theory with
$\alpha_0=10^{-4}$ and $\beta_0=-4.35$. A polytropic core is collapsed
using the fiducial hybrid EOS for three different resolutions (see text for
details). The top panels show the evolution of the mass density $\rho$
(left) and the scalar-field $\varphi$ (right) for the highest-resolution
run a $t=7$ (grey), $28$ (yellow), $37$ (green), $38$ (blue), $57$
(red) ms after starting the simulations. Bounce happens at $t \sim
38~{\rm ms}$ and the shock reaches the surface of the star at
$t\sim 131~{\rm ms}$.
The bottom panels show the self-convergence properties of the
gravitational mass $m$ (left) and the scalar field $\varphi$ (right)
at the same times. As detailed in the text, solid and dotted
(dashed) lines are expected to coincide for second- (first-) order
convergence. We initially observe second order convergence which
decreases to first order due to (i) the shock capturing scheme
when a discontinuity forms at bounce and (ii) numerical noise
in the scalar field propagating in from the outer boundary.
}
\label{convtest}
\end{figure}

Here we present the convergence properties of our dynamical code.  Given
three simulations of increasing resolutions with grid spacings
$\Delta r_1>\Delta r_2>\Delta r_3$,
the self-convergence factor $Q$ of a quantity $q$ is defined by
\begin{align}
Q= \frac{q_1-q_2}{q_2-q_3} =  \frac{(\Delta r_1)^n
        - (\Delta r_2)^n}{(\Delta r_2)^n - (\Delta r_3)^n}\,,
\end{align}
where $q_i$ indicates the quantity $q$ obtained at resolution
$\Delta r_i$ and $n$ is the convergence order of the implemented
numerical scheme.

We collapse a $\Gamma=4/3$ polytropic core in ST theory with
$\alpha_0=10^{-4}$ and $\beta_0=-4.35$ using the hybrid EOS with
$\Gamma_1=1.3$, $\Gamma_2=2.5$ and $\Gamma_{\rm th}=1.35$.
This model is evolved for
three uniformly spaced\footnote{For the convergence analysis,
we use uniform grids exclusively, i.e.~do not switch to
a logarithmic spacing in the outer parts. Non-linear grid
structure would make a quantitative convergence analysis
highly complicated.}
grids of size $R_{\rm out} =
2\times 10^3 ~{\rm km}$ with $N=6000,12000$ and
$24000$ grid cells. For these grids, we
expect $Q=2$ ($Q=4$) for first- (second-)
order convergence. The bottom panels of Fig.~\ref{convtest} show
the convergence properties of the gravitational mass $m$ and the
scalar field $\varphi$ at various timesteps. Solid lines show the
difference between the coarse and the medium resolution runs
$q_1-q_2$; dashed (dotted) lines show the difference between the
medium and the fine resolution runs $q_2-q_3$ multiplied by the
expected first- (second-) order self-convergence factor  $Q=2$
($Q=4$). Second-order convergence is achieved if the solid
and dotted lines coincide, while the code is only first-order
convergent if the solid and dashed lines coincide. The evolution of
$\rho$ and $\varphi$ is displayed in the top panels for orientation.

The enclosed gravitational
mass $m$ shows good second-order convergence properties
before bounce $t\lesssim 38~{\rm ms}$, while  convergence  deteriorates
to first order as the shock propagates outwards at $t\gtrsim 38~{\rm
ms}$. This is a characteristic feature of high-resolution shock-capturing
schemes; they are second-order (or higher) schemes for smooth fields,
but become first-order accurate
in the presence of discontinuities
\cite{2010CQGra..27k4103O}.
Note that the behaviour of the total gravitational and baryonic
mass is more complex than in the GR limit where both are conserved
because of the absence of gravitational radiation in spherical
symmetry and the vanishing of the source term $s_D$ in Eq.~(\ref{sourceD}).
The convergence properties of the scalar field are more complicated.
While evolved with a second-order accurate scheme, we observe that
the scalar field's
convergence may deteriorate for the following two reasons: (i) the
drop to first-order convergence of the matter fields which source
the scalar dynamics; (ii) numerical noise generated at the outer
boundary, especially during the early transient (note that in this
convergence analysis the outer boundary is located much closer to the
core than in our production runs because of the limit imposed by
a strictly uniform grid). The observed
convergence in the scalar field bears out these effects. Initially
convergent at second order, we note a drop to roughly first order
after one light crossing time $R_{\rm out}/c\sim 7~{\rm ms}$.
As the noise is gradually dissipated away, the convergence increases
back towards second order, but drops once more to first order at
the time the shock forms in the matter profile around $38~{\rm ms}$.

We also tested the convergence of the scalar waveform $\varphi(t)$
extracted at finite radius in these simulations and observe
first-order convergence which we attribute to the relatively small
total computational domain such that the outer boundary effects
discussed above causally affect the extraction radius early in the
simulation. The resulting uncertainty in the waveform is obtained by
comparing the finite resolution result with the Richardson
extrapolated (see, e.g.,~\cite{Kreiss1973}) waveform. We find a
relative error of $10\%$ which we regard as a conservative estimate
since the production runs 
have
much larger computational domains.

\section{Results and discussion}
\label{results}
In this Section, we present the results of our simulations. After illustrating
the main features of stellar collapse in ST theories (Sec.~\ref{ccdyn}),
we present our predictions for monopole gravitational radiation
(Sec.~\ref{monopolegw}).  All waveforms presented in this section are publicly available at \cite{DGwebsite}. 

\subsection{Core-collapse dynamics}
\label{ccdyn}

\begin{figure}
\includegraphics[width=\textwidth]{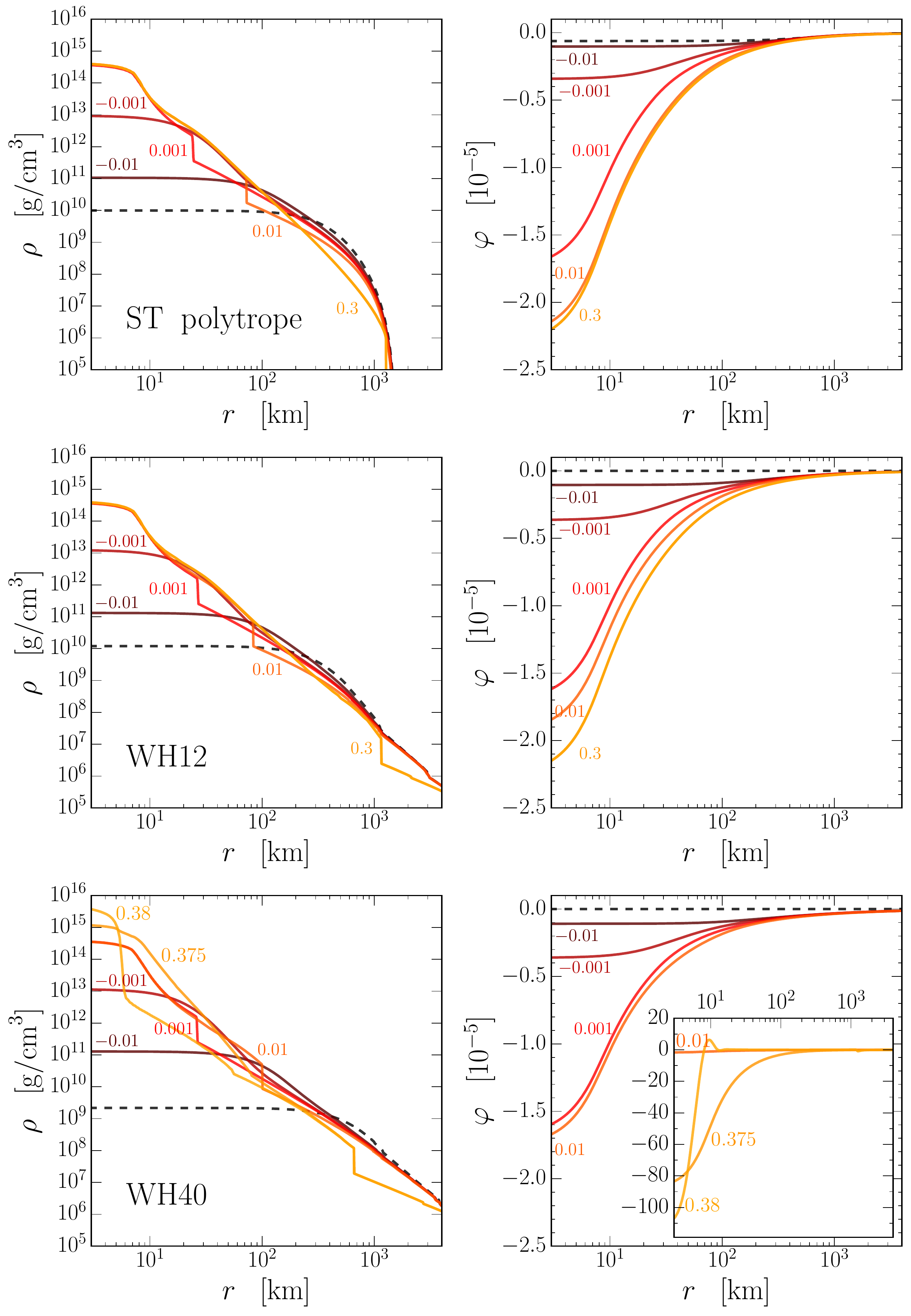}
\caption{
 Collapse of a $\Gamma=4/3$ polytrope (top), the $12 M_{\odot}$ (centre) and $40 M_{\odot}$ (bottom)
  pre-SN profiles of  \cite{2007PhR...442..269W} 
 in ST theory with $\alpha_0=10^{-4}$  and $\beta_0=-4.35$,
 assuming $\Gamma_1=1.3$,
 $\Gamma_2=2.5$ and $\Gamma_{\rm th}=1.35$.
 The evolution of the mass density $\rho$ (left) and the scalar
 field $\varphi$ (right) is shown as a function of the radius $r$
 at various timesteps $t-t_{B}=-0.01,-0.001,0.001,0.1, 0.3, 0.375,0.38$ s, measured from the bounce time $t_B$. 
 Timesteps are coloured from darker
 (early times) to lighter (late times) solid lines as labelled;  initial profiles are shown with  black dashed lines. The inset in the bottom right panel shows the wide variation of the scalar field when a BH is formed. An animated version of this figure is available online at Ref.~\cite{DGwebsite}. 
 }
\label{evolution}
\end{figure}

\begin{figure}
\centering
\includegraphics[width=1\textwidth]{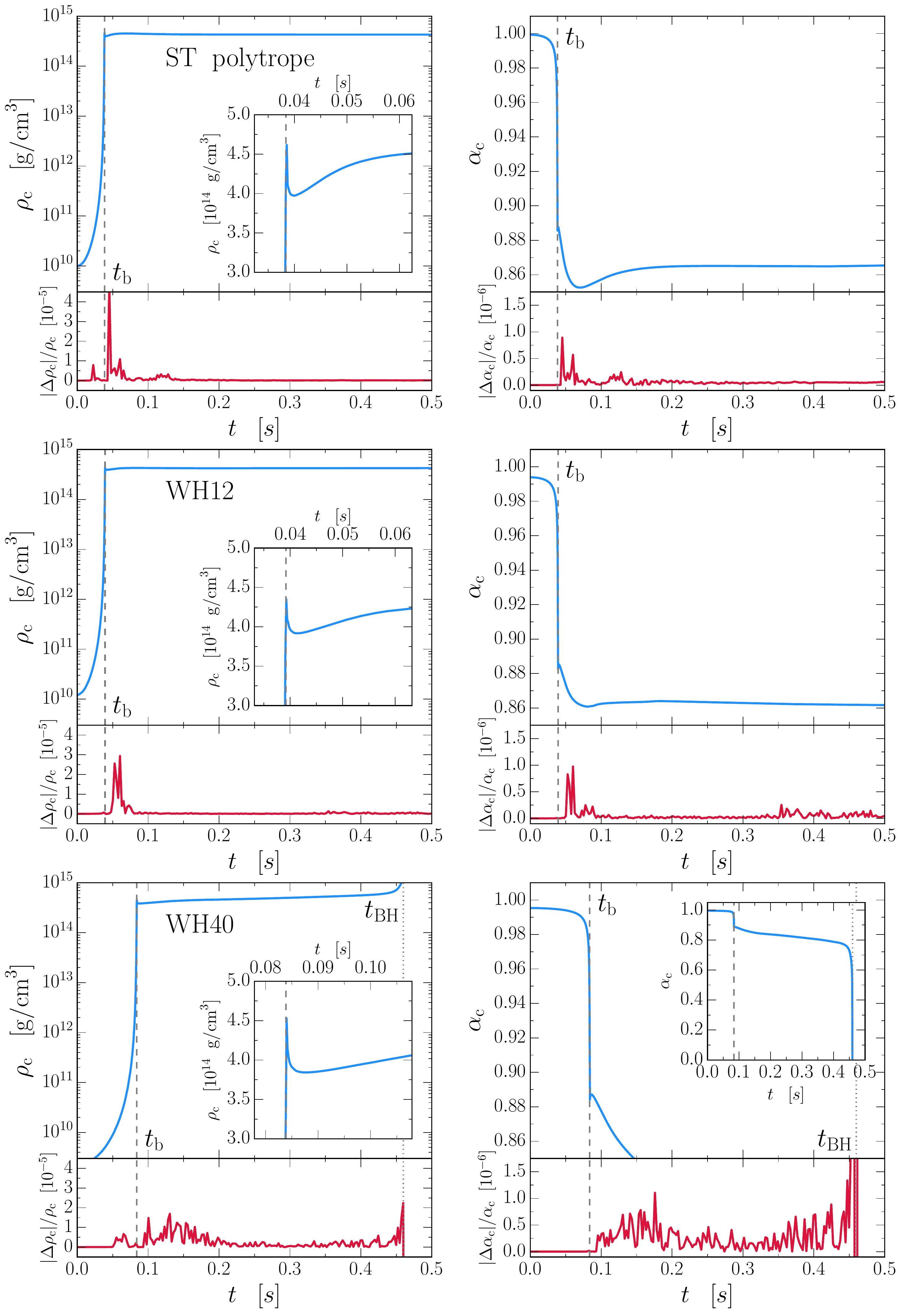}
\caption{Evolution of the central values of the mass density $\rho_c$ (left panels) and lapse function $\alpha_c$ (right panel) through collapse, bounce and late time evolution in ST theory with $\alpha_0=10^{-4}$
 and $\beta_0=-4.35$. We use the hybrid EOS with fiducial parameters ($\Gamma_1=1.3$, $\Gamma_2=2.5$ and $\Gamma_{\rm th}=1.35$) and three different initial profiles: ST polytrope (top), WH12 (centre) and WH40 (bottom).
 Gray dashed lines mark the bounce time $t_b$; the WH40 profile first collapses to a protoneutron star and then to a BH at $t_{\rm BH}\sim 0.46\,\mathrm{s}$ marked by grey dotted lines. Relative differences with  analogous simulations performed in GR are shown in the lower subpanels (red lines). Deviations 
in the dynamics are very small: of the order of $|\Delta\rho_c|/\rho_c\sim 10^{-5}$ and $|\Delta\alpha_c|/\alpha_c\sim 10^{-6}$.}
\label{centralvalues}
\end{figure}

The main features of the core-collapse dynamics are summarised in
Figs.~\ref{evolution} and \ref{centralvalues}. We present the collapse
of both a polytropic core and two realistic pre-SN models
(Sec.~\ref{indata}) in ST theory with $\alpha_0=10^{-4}$ and
$\beta_0=-4.35$. 
These parameter choices lie
on the edge of the parameter space region compatible with binary
pulsar experiments (cf. Fig.~\ref{obsconstraints}) and marginally
allow  for spontaneous scalarisation
\cite{1998PhRvD..58f4019N,1998PhRvD..57.4802H}. Collapse is performed
using the hybrid EOS (Sec.~\ref{EOS}) with fiducial values
$\Gamma_1=1.3$, $\Gamma_2=2.5$ and $\Gamma_{\rm th}=1.35$.

Since $\Gamma_1<4/3$, the initial iron cores are not equilibrium
solutions of the evolution equations and collapse is triggered
dynamically.  While the polytropic profile collapses
smoothly from the very beginning of the simulation, a brief transient
in the scalar-field evolution is present in the early stages of the
collapse of both the WH12 and WH40 models.  As already mentioned in
Sec.~\ref{indata}, this is due to the fact that these initial models
are Newtonian and their initially vanishing scalar profiles are not
fully consistent with the ST theory used in the evolution.  This
transient generates a pulse in the scalar field propagating outwards
at the speed of light.
The scalar field quickly settles down in the stellar interior while the spurious pulse
reaches the outer edge of the grid at $R_{\rm out}/c\sim 0.6 ~{\rm s}$ where it is absorbed by the
outgoing boundary condition.

As the collapse proceeds in either of the three models, the central
mass density increases from its initial value up to beyond nuclear
densities $\rho_{\rm nuc} \simeq 2 \times 10^{14} ~{\rm g\,cm}^{-3}$.
The EOS suddenly stiffens to an effective adiabatic index
$\Gamma_2\gg4/3$ and the inner core bounces
after $t_{b} \sim$ 38, 39, and 84~ms from the
beginning of the simulations, for the ST polytrope, WH12 and WH40
profile, respectively\footnote{The WH40 profile takes longer to reach
  $\rho_{\rm nuc}$ because of its lower initial central density
  (cf. Fig.~\ref{initialprofiles}).}. 
Core bounce
  launches a hydrodynamic shock into the outer core. Due to the steep
  density profile of the polytrope, the shock explodes the polytrope
  promptly, reaching its surface at $\sim 130 ~{\rm ms}$ from the
  start of the simulation. Since we set $\Gamma_\mathrm{th} = 1.35$ to
  qualitatively account for reduced pressure due to nuclear
  dissociation and neutrino losses, the pressure behind the shock is
  not sufficient to lead to a prompt explosion of the more realistic
  WH12 and WH40 progenitors. The shock stalls and only secularly moves
  to larger radii as the accretion rate decreases.
Core bounce is paralleled by
a small reversal in the scalar field amplitude.
For example, in the collapse of the polytropic model shown in Fig.~\ref{evolution}, the central value of $\varphi$ 
reaches a minimum $\sim -2.6\times
10^{-5}$ at bounce before settling down to $\sim-2.3\times 10^{-5}$. 
A more detailed description of the scalar field dynamics is
postponed to Sec.~\ref{monopolegw}.

The inner regions of the 
promptly exploding
  polytropic model settle down to a stable compact remnant with
  compactness $m/r\sim0.053$ (measured from the metric potential at
  $r=10~{\rm km}$). While simulations with model WH12 show that the
  shock stalls and then only slowly shifts to larger radii, the low
  accretion rate in this model does not increase its compactness above
  the values that we find for the polytropic model. In both models,
the scalar charge $\omega$ evolves from $\sim- 10^{-4}~M_\odot$ to
$\sim-2\times 10^{-4}~M_\odot$ during the
entire evolution
and thus
remains of the order of $\omega/M\sim \alpha_0$, as predicted for
weakly scalarised NS solutions (cf.\ Sec.~\ref{coupfun}).} In both
simulations, the NSs do not evolve to strongly spontaneously
scalarised solutions because the compactness of the core remains lower
than the threshold at which multiple solutions appear ($m/r\sim
0.2$\cite{2004AIPC..736...35E}).

On the other hand, the WH40 
model forms a protoneutron star that subsequently
  collapses to a BH within
$t_{\rm BH}\sim0.46\,\mathrm{s}$ from the beginning of the simulation
($\sim 0.38\,\mathrm{s}$ from bounce). 
The high  accretion rate in this model
    quickly increases the central
compactness. As BH formation is approached, our gauge choice causes
the lapse function $\alpha$ to collapse to zero near the origin
(Fig.~\ref{centralvalues}) and the dynamics of the inner region
effectively freezes. In this regime, spontaneously scalarised NS
solutions are not only present but energetically favourable
\cite{1993PhRvL..70.2220D,1996PhRvD..54.1474D,1997PThPh..98..359H}. While collapsing
towards a BH, the core first transits through a spontaneously
scalarised NS. BH formation generates strong scalar-field excitation,
enhanced in this case ($\beta_0=-4.35$) by spontaneous scalarisation
(cf. Sec.~\ref{STeffect}). The central value $\varphi_c$, which
through collapse and bounce remains close to values of the order of
$\sim - 10^{-5}$, increases in magnitude to $\sim - 2 \times
10^{-3}$. This signal propagates outwards at
the speed of light, rapidly leaves the region of the stalled shock,
and reaches the extraction radius after about $\sim 0.56$ s from the
beginning of the simulation.

Our gauge choice does not allow us to follow the evolution of the inner
region of the star beyond BH formation. Following
\cite{1998PhRvD..57.4789N}, we terminate the evolution of the matter
variables at the onset of BH formation in order to ensure numerical
stability. At this point, the inner core has reached a compactness of
$\sim 0.466$, close to the BH value of $0.5$. We are still able to
gain insight into the late-time behaviour of the scalar field,
however, by solving the wave equation (\ref{jordanevolphi}) on the now
frozen background (cf. Fig.~\ref{evolution}). We observe in these
evolutions that, as the NS (now spontaneously scalarised) collapses to
a BH, the scalar field slowly relaxes to a flat profile as predicted
by the no-hair theorems
\cite{1972CMaPh..25..167H,1971ApJ...166L..35T,Chase:1970,1997siad.conf..216B,2012LRR....15....7C}.

Overall, the entire dynamics strongly resembles GR. The scalar field
is mostly driven by the matter evolution, which in turn is largely
independent of the scalar field propagation. This point is illustrated
in Fig.~\ref{centralvalues}, where the central values of the mass
density $\rho_c$ and lapse function $\alpha_c$  obtained in ST
theory and in GR are compared. The relative differences between these
two scenarios are about $|\Delta\rho_c|\rho_c\sim 10^{-5}$ and
$|\Delta\alpha_c|/\alpha_c\sim 10^{-6}$ throughout collapse, bounce,
and (eventually) BH formation.

\subsection{Monopole gravitational-wave emission}
\label{monopolegw}
Unlike GR, ST theories of gravity admit propagating monopole GWs. This
breathing mode of the scalar field is potentially detectable with
current and future GW interferometers which have therefore the
potential of constraining the parameters of the theory.  We now
analyse the scalar 
 GW signal extracted from our
numerical simulations, separately discussing the effects of the EOS
and the ST parameters.  
 \subsubsection{Effect of the equation of state}
 
\label{EOSeffect}

 \begin{figure}\centering
\includegraphics[width=\textwidth]{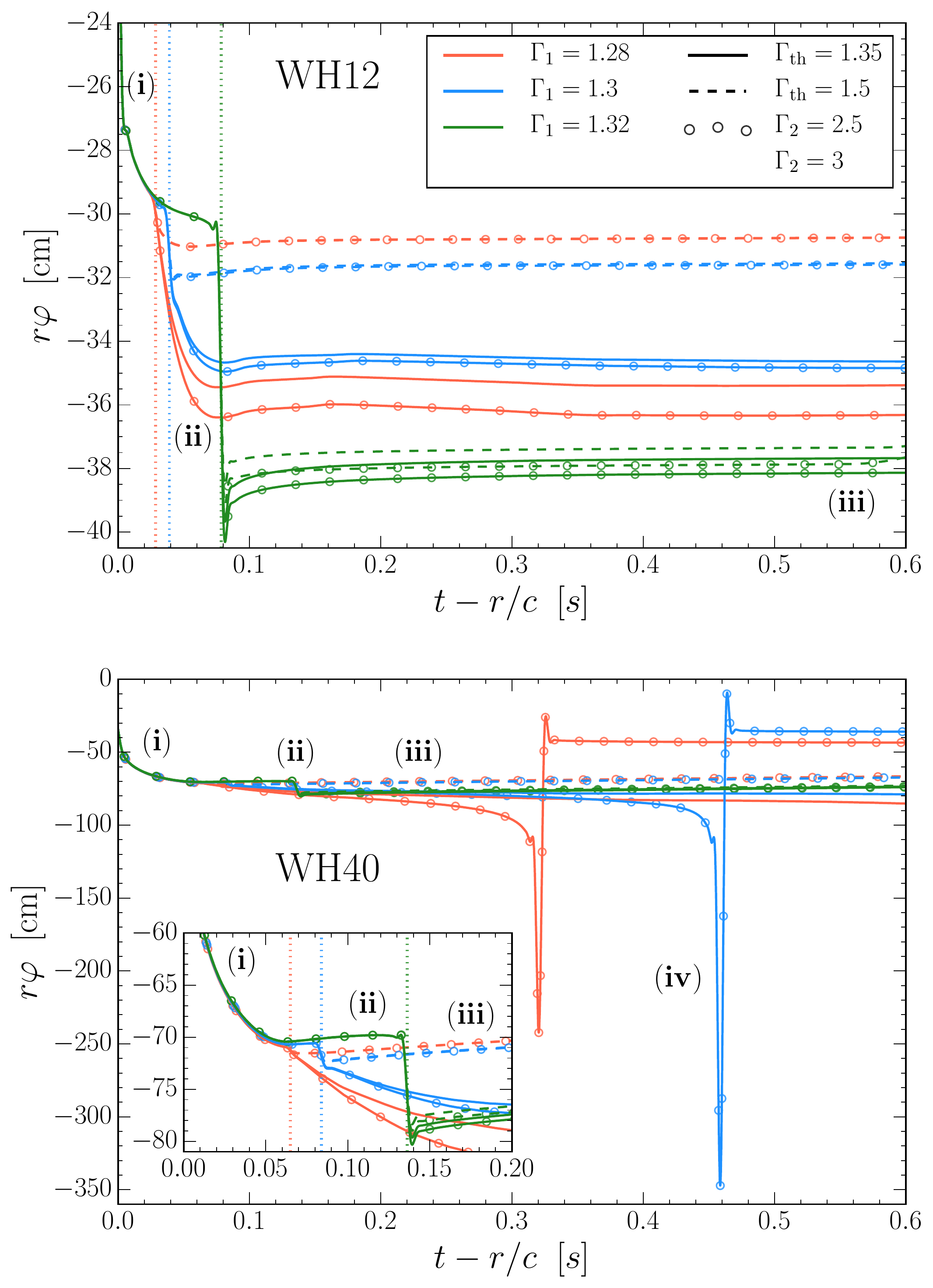}
\caption{ Effect of the hybrid EOS adiabatic indices on the emitted
  monopole gravitational waveform $h(t)\propto r \varphi$
  [cf. Eq.(\ref{monopoleht})]. The signal is plotted against the
  retarded time $t-r/c$ at the extraction radius. Simulations are
  performed using the preSN initial profiles WH12 (top) and WH40
  (bottom) in ST theory with $\alpha_0=10^{-4}$ and
  $\beta_0=-4.35$. The curves encode the value of $\Gamma_1$ in their
  brightness (colour) and the value of $\Gamma_{\rm th}$ in their
  line style: $\Gamma_1=1.28$ (red), $1.3$ (blue), $1.35$ (green);
  $\Gamma_{\rm th}=1.35$ (solid), $1.5$ (dashed).  For each of these
  combinations, two curves are present: circles mark simulations with
  $\Gamma_2=2.5$, while no symbols are shown for $\Gamma_2=3$. For
  some cases, these two curves overlap to such high precision that
  they become indistinguishable in the plot. The lower-case Roman
  labels refer to the key phases of the GW signal described in
  Sec.~\ref{EOSeffect}: {\bf (i)} initial pulse of the spurious
  radiation; {\bf (ii)} collapse and bounce; {\bf (iii)} NS
  configuration; {\bf (iv)} BH formation. The bounce time is marked
  with vertical dotted lines following the same colour codes of the
  other curves.  Note that $\Gamma_1$ is the only adiabatic index that
  has an effect on the bounce time. Waveforms presented in this figure are available at \cite{DGwebsite}. }
\label{rphi}
\end{figure}

As detailed in Sec.~\ref{EOS}, the hybrid EOS is a simplified EOS
model that qualitatively approximates more sophisticated microphysical
EOS in the core collapse context (e.g.,
\cite{1991NuPhA.535..331L}). The hybrid EOS is characterised by three
adiabatic indices for the pre-bounce dynamics ($\Gamma_1$), the
repulsion at nuclear densities ($\Gamma_2$), and the response of the
shocked material ($\Gamma_{\rm th}$).  The effect of the EOS on the
emitted GW waveforms is explored in Fig.~\ref{rphi}, where we show
time-domain monopole waveforms $h(t)\propto r \varphi$ [cf.
  Eq.~(\ref{monopoleht}) with $\varphi_0=0$] for various choices of
$\Gamma_1$, $\Gamma_2$, and $\Gamma_{\rm th}$. All simulations shown in
Fig.~\ref{rphi} are performed in ST theory with $\alpha_0=10^{-4}$ and
$\beta_0=-4.35$, the lower limit of $\beta_0$ compatible with present
observations, using the WH12 and WH40 initial profiles. We plot the GW
signals as a function of the retarded time $t-r/c$, such that the
origin corresponds to a single light-crossing time at the extraction
radius $r_{\rm ext}=3\times 10^4~{\rm km}$.

The structure of the emitted signals displayed in
Fig.~\ref{rphi} consists of the following four main stages.
\begin{enumerate}
\item
The initial pulse of spurious radiation arises from the initialisation
of the scalar field, as already pointed out in Secs.~\ref{indata}
and \ref{ccdyn}. This pulse propagates outwards and leaves the
extraction region after a retarded time of
about $0.006$~s.
\item 
As the core collapses, the scalar field signal significantly grows in
amplitude.  Although the first $\sim0.02$ s of the waveform appear to
be rather insensitive to the EOS, the adiabatic indices strongly
affect the total amount of time the star spends in the collapse phase
before bounce. In the hybrid EOS, this is controlled by $\Gamma_1$.
Collapse is triggered by $\Gamma_1 \le 4/3$ and the smaller $\Gamma_1$,
the more rapid the collapse and the smaller the mass of the inner core that
collapses homologously (e.g., \cite{1997A&A...320..209Z,2007PhRvL..98y1101D}).
We note that in reality (and in simulations using more realistic
microphysics), $\Gamma_1$ is not a parameter. Instead, the effective
adiabatic index is a complex function of the thermodynamics and
electron capture during collapse
\cite{2007PhRvL..98y1101D,2008PhRvD..78f4056D}.
Figure~\ref{rphi} shows that core bounce occurs in our simulations at
retarded time $t-r_{\rm ext}/c \sim 0.03, 0.04, 0.08$ s
($0.06,0.07,0.14$) for model 
WH12 (WH40) and
$\Gamma_1=1.28,1.3,1.32$ respectively.  The bounce itself is a rapid
process with a duration of $\Delta t \sim 1-2$ ms.

\item
After bounce, the scalar field in the inner core settles down to a
non-trivial profile, as illustrated in Fig.~\ref{evolution}. The
post-bounce value of $\varphi$ at $r_{\rm ext}$, hence the value of
$h(t)$, encodes information about all three adiabatic indices. In
particular, larger values of $\Gamma_1$ and smaller values of
$\Gamma_{\rm th}$ both produce stronger wave signals $h(t)$, which can
be intuitively understood as follows.  Larger values of $\Gamma_1$
result in a more massive inner core in the pre-bounce stage because
the speed of sound is larger and, hence, more matter remains in sonic
contact in the central region.
At bounce, this implies a more compact core and a
correspondingly larger amplitude in the scalar wave.  Smaller values
of $\Gamma_{\rm th}$ imply lower pressure in the shocked material,
and, therefore, material that accretes through the shock settles
faster onto the protoneutron star. In terms of microphysical
processes, this effect is driven by neutrino cooling
\cite{2007PhRvL..98y1101D,2008PhRvD..78f4056D}, which is not included
in our simulations.  In contrast, we find that $\Gamma_2$ has a
relatively minor effect on this phase of the wave signal: the scalar
field profile is only slightly more pronounced for lower values of
$\Gamma_2$, which result in a deeper bounce and a more compact
postbounce configuration. Note that in the waveforms shown in
Fig.~\ref{rphi}, both $\Gamma_{\rm th}$ and $\Gamma_2$ only affect the
wave signal at and after bounce.  This is expected, since
$\Gamma_{\rm th}$ only plays a role in the presence of shocked material
and $\Gamma_2$ affects only the high density regime of the EOS not
encountered in the collapse evolution prior to bounce.

\item Two of the simulations shown in Fig.~\ref{rphi} (namely
  $\Gamma_1=1.28,1.3$; $\Gamma_2=2.5$; $\Gamma_{\rm th}=1.35$) for the
  WH40 profile collapse to BHs. BH formation is triggered when the
  protoneutron star exceeds its maximum mass and is therefore
  facilitated and accelerated by smaller values of the adiabatic
  indices.  BH formation generates a very large pulse in the scalar
  field which dominates the entire GW signal. Spontaneous
  scalarisation (marginally allowed for the value $\beta_0=-4.35$
  chosen here) before BH formation further enhances the signal. The
  amplitude of the scalar field signal from this phase is more than an
  order of magnitude larger than the bounce signal in the absence of
  BH formation. We expect BH-forming collapse events to be the most
  promising source of monopole GWs  in the context of ST theory.
\end{enumerate}

\subsubsection{Effect of the ST parameters.}
\label{STeffect}

\begin{figure}\centering
\includegraphics[width=\textwidth]{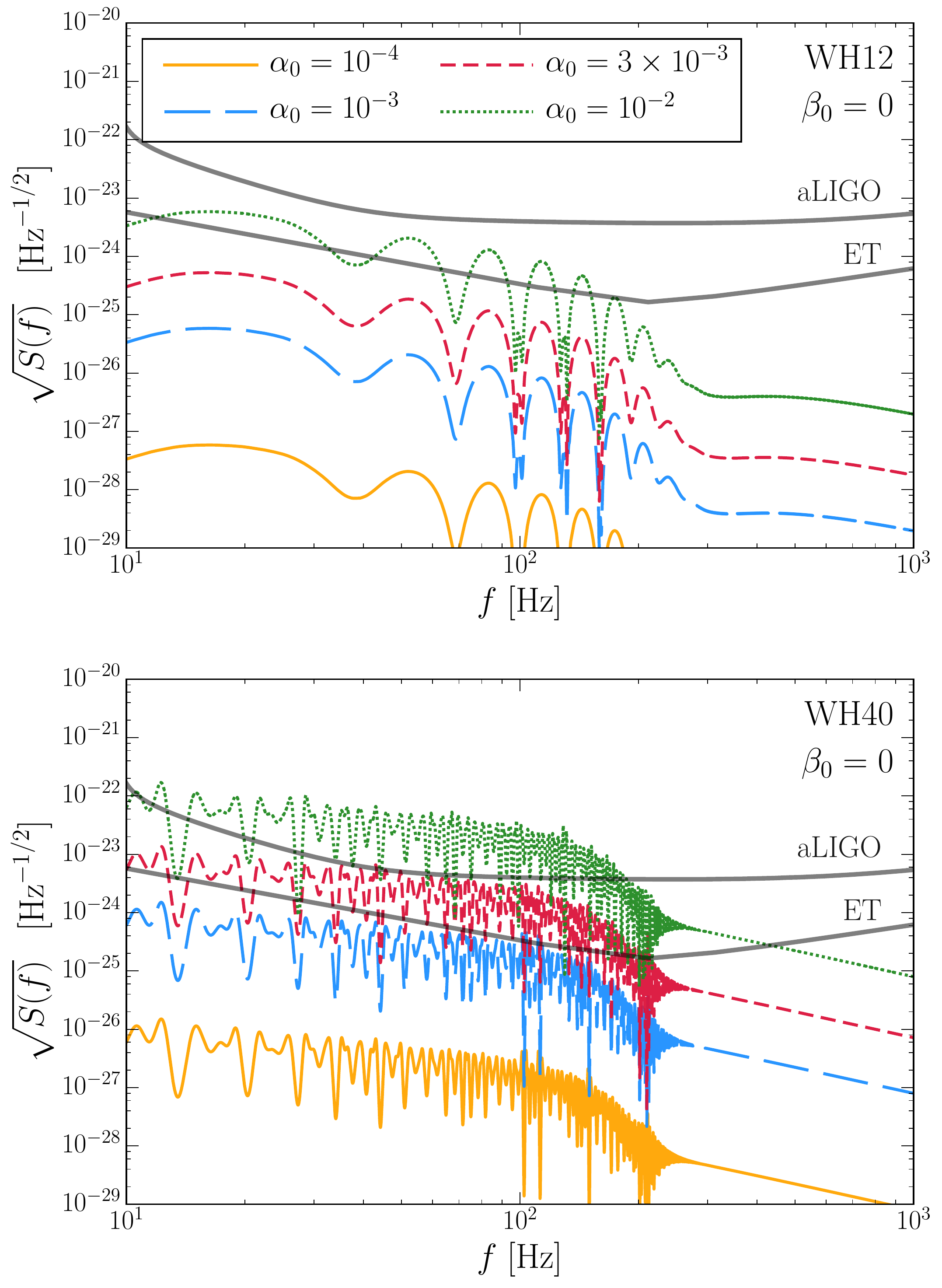}
\caption{Effect of $\alpha_0$ on frequency domain waveforms for monopole GWs emitted during stellar collapse. The $M_{\rm ZAMS}=12 M_{\odot}$ (top) and $40 M_{\odot}$ (bottom) pre-SN models  of  Ref.~\cite{2007PhR...442..269W} are evolved using the hybrid EOS with $\Gamma_1=1.3$, $\Gamma_2=2.5$ and $\Gamma_{\rm th}=1.35$. 
Four simulations are presented for fixed $\beta_0=0$ (equivalent to Brans-Dicke theory): $\alpha_0=10^{-4}$ (orange, solid), $10^{-3}$ (blue, long-dashed), $3\times 10^{-3}$ (red, short-dashed), $10^{-2}$ (green, dotted). These values are compared with current experimental constraints in Fig.~\ref{obsconstraints}.
We consider optimally oriented sources placed at $D=10$ kpc and compare them with the expected sensitivity curves of Advanced LIGO and the Einstein Telescope. Waveforms presented in this figure are available at \cite{DGwebsite}. }
\label{spanSTa0}
\end{figure}

\begin{figure}\centering
\includegraphics[width=\textwidth]{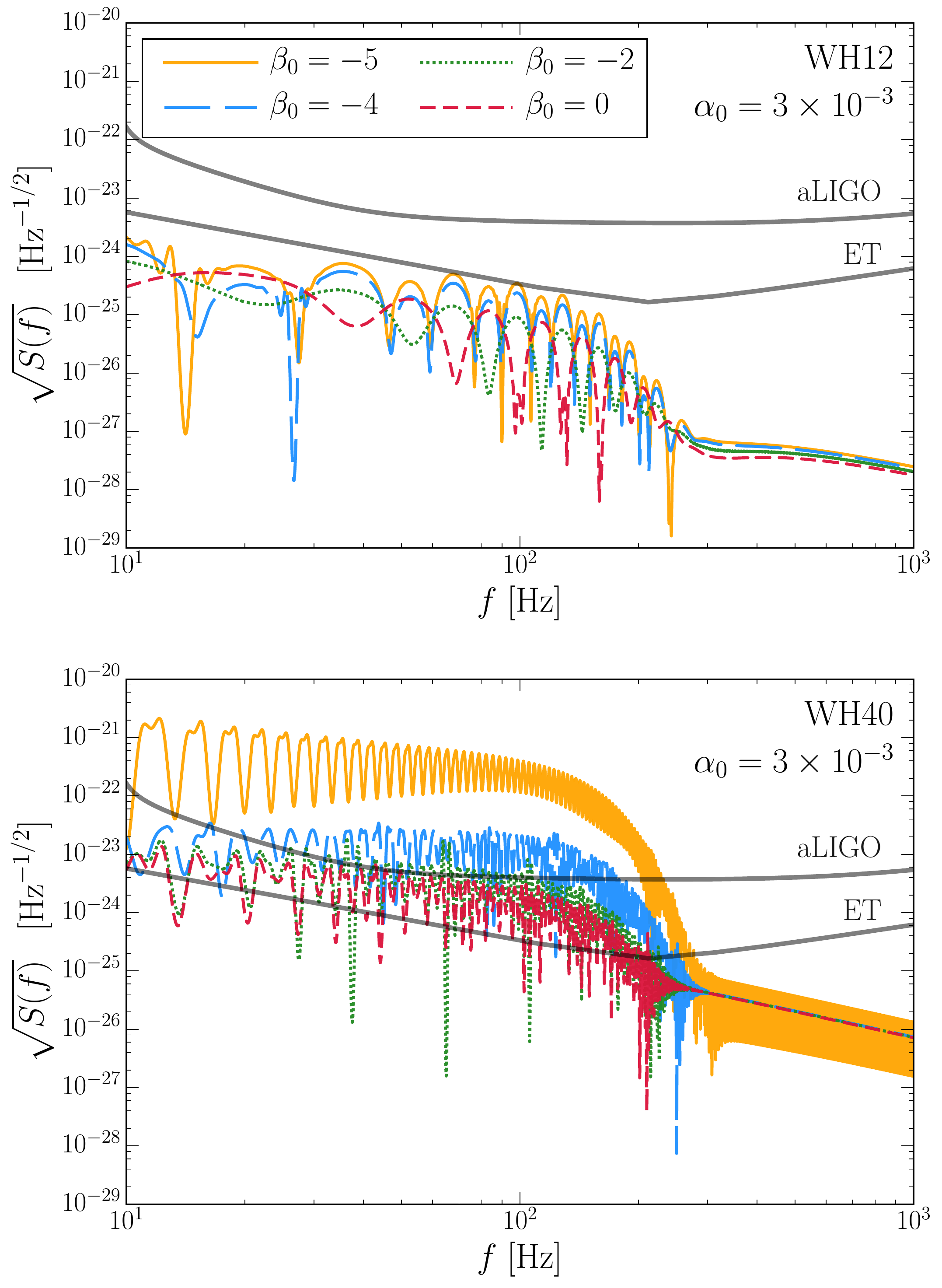}
\caption{Effect of $\beta_0$ on frequency domain waveforms for monopole GWs emitted during stellar collapse. The $M_{\rm ZAMS}=12 M_{\odot}$ (top) and $40 M_{\odot}$ (bottom) pre-SN models  of  Ref.~\cite{2007PhR...442..269W} are evolved using the hybrid EOS with $\Gamma_1=1.3$, $\Gamma_2=2.5$ and $\Gamma_{\rm th}=1.35$. 
Four simulations are presented for fixed $\alpha_0=3\times 10^{-3}$ (marginally allowed by solar-system constraints): $\beta_0=-5$ (orange, solid), $-4$ (blue, long-dashed), $-2$ (green, dotted), $0$ (red, short-dashed). These values are compared with current experimental constraints in Fig.~\ref{obsconstraints}.
We consider optimally oriented sources placed at $D=10$ kpc and compare them  with the expected sensitivity curves of Advanced LIGO and the Einstein Telescope.  Waveforms presented in this figure are available at \cite{DGwebsite}. 
}
\label{spanSTb0}
\end{figure}

As introduced in Sec.~\ref{coupfun}, PN deviations from GR in ST
theories only depend on two parameters, $\alpha_0$ and $\beta_0$.
While $\alpha_0$ mainly controls the perturbative deviation from GR,
$\beta_0$ is responsible for non-linear effects such as spontaneous
scalarisation.  Our primary interest in this section is to explore the
effect of these parameters on the detectability of signals with
current and future GW detectors and, in particular, comparison with their
sensitivity curves.

Figures \ref{spanSTa0} and \ref{spanSTb0} show frequency domain
waveforms $\sqrt{S_h(f)}$ compared with the expected (design)
sensitivity curves $\sqrt{S_n(f)}$ of Advanced LIGO and the Einstein
Telescope. We use the WH12 and WH40 initial profiles, together with
the hybrid EOS with fiducial values $\Gamma_1=1.3$, $\Gamma_2=2.5$, and
$\Gamma_{\rm th}=1.35$ (cf.\ Sec.~\ref{EOS}).  To better disentangle
the effect of the two ST parameters, Fig.~\ref{spanSTa0}
(\ref{spanSTb0}) presents a series of simulations where only
$\alpha_0$ ($\beta_0$) varies while the other parameter is kept fixed
at $\beta_0=0$ ($\alpha_0=3\times 10^{-3}$). These two parameter sets
overlap at $\alpha_0=3\times 10^{-3}$ and $\beta_0=0$ and this
specific simulation is shown in both figures. The location of our runs
in the \mbox{$(\alpha_0,\beta_0)$} parameter space is shown in
Fig.~\ref{obsconstraints}.
Throughout our analysis, we consider optimally oriented sources placed at a fiducial distance
of $D=10$ kpc, i.e.~within the Milky Way.

As mentioned above, the most pronounced feature in the emitted
waveform arises from the collapse of the protoneutron star to a BH. As
a consequence, the GW strains emitted during collapse of the $M_{\rm
  ZAMS}=40 M_{\odot}$ profile WH40 are over an order of magnitude
larger than the corresponding signals obtained from the collapse of
the WH12 profile. BH formation (possibly enhanced by spontaneous
scalarisation -- see below) following the
protoneutron star phase
    is the most promising signature of monopole GWs in the
context of ST theory.

Simulations presented in Fig.~\ref{spanSTa0} are performed in ST
theory with $\beta_0=0$ and various values of $\alpha_0$, equivalent
to Brans-Dicke theory with $\omega_{\rm BD}=
(1-6\alpha_0^2)/2\alpha_0^2$. Since spontaneously scalarised stars are
not permitted in this regime, this set of simulations illustrates the
effect of perturbative deviations from GR.
In ST theory with $\alpha_0\sim 3\times 10^{-3}$, just compatible with
the Cassini bound, GW signals generated by BH formation in our
  Galaxy, are
marginally detectable by second-generation ground-based detectors
and fall well into the sensitivity range of future experiments like the
Einstein Telescope.
Observation of a BH forming core collapse event with Advanced LIGO
therefore has the potential of independently constraining
ST theory at a level comparable with the most stringent present tests.
Future third-generation observatories, on the other hand,
will be able to push the constraint to new levels: $\alpha_0 \lesssim \times 10^{-4}$ corresponding to $|\gamma^{\rm PPN} -1|\lesssim 2 \times 10^{-8}$;
cf.~Eq.~(\ref{eddppn}). On the other hand, our present results suggest that
core collapse forming NSs (such as in our WH12 model) will at best
allow for an independent confirmation of existing bounds, even when
observed with third-generation observatories.

By analysing the curves in Fig.~\ref{spanSTa0} quantitatively, we observe
that the amplitude of the GW signal scales approximately as
$\alpha_0^2$. One factor of $\alpha_0$ is evidently due to the
local coupling between the scalar field and the detector [see
Eq.~(\ref{monopoleht})]. In our simulations, however, we find
that the amplitude of the emitted scalar field $\varphi$ also depends (roughly linearly) on $\alpha_0$.
This second factor of $\alpha_0$ is entirely due to the source dynamics
and therefore separate from that arising in
the coupling between the wave and the detector at the moment of observation.
Even though the dynamics in the matter variables only mildly deviates
from the GR case (cf. Fig.~\ref{centralvalues}), such perturbative deviations from GR of the
order $\alpha_0$
can leave a significant imprint on the generation of monopole GWs.

The strongest effect of $\beta_0$ on the structure of NSs is that of allowing for spontaneously scalarised stars in the range $\beta_0 \lesssim -4.35$. In fact, it is precisely the strength of this effect that enables binary pulsar observations to severely constrain $\beta_0$ as displayed in Fig.~\ref{obsconstraints}. For our simulations using values of $\beta_0$ significantly above
the spontaneous scalarisation threshold of about $-4.35$,
we only identify a relatively minor variation of the
scalar wave with $\beta_0$. This is well illustrated by the curves in Fig.~\ref{spanSTb0} corresponding
to $\beta_0=0$ and $-2$.
Deviations of this kind can become particularly pronounced
for $\beta_0\lesssim-4.35$ {\em if} the strongly non-linear effects
of spontaneous scalarisation are triggered
\cite{1998PhRvD..58f4019N,1998PhRvD..57.4802H,1993PhRvL..70.2220D,1996PhRvD..54.1474D}; cf.~Fig.~\ref{spanSTb0}. Whether this effect is triggered in
our simulations and, in consequence,
the shape and magnitude of the resulting waveform,
critically depends on the stellar progenitor.

\begin{itemize}

\item If the core
  collapse leads to a protoneutron star that subsequently collapses to
  a BH, spontaneous scalarisation can be triggered  by the high compactness reached shortly before BH formation, leading to a large
  enhancement of the GW signal. In the bottom panel of
  Fig.~\ref{spanSTb0}, we compare the frequency-domain GW signals for
  the BH-forming WH40 progenitor. Spontaneous scalarisation occurs in
  a very strong way for the model with $\beta_0 = -5$ (already ruled
  out by current constraints) and leads to an enhancement of $\sim$two
  orders of magnitude in the amplitude compared to models that do not
  exhibit this strong non-linear behaviour (cases with $\beta_0 = 0$
  and $\beta_0 = -2$). The waveform of the model with $\beta = -4$
  (still allowed) is also somewhat enhanced by non-linear scalar field
  dynamics. Given the quantitative differences between these
  waveforms, present and future detectors have the potential of either
  observing scalar waves from BH-forming core collapse events or use
  their absence in the data stream to constrain the parameter
  $\beta_0$ beyond current limits. This will, however, require that
  other uncertain parameters such as the distance to the source etc.
  can be determined with high precision.
\item
None of our simulations of the progenitor model WH12 leads to BH
formation in the time simulated. This is so because this moderate-mass
progenitor has a steep density gradient outside its core and thus a
lower postbounce accretion rate. If no explosion is launched, a BH
would still result, though on a timescale of
$\mathcal{O}(10)\,\mathrm{s}$ \cite{2011ApJ...730...70O}. Furthermore,
we do not observe any signature of spontaneous scalarisation in the
waveform or 
in
  the protoneutron star of the WH12 model,
  even for
the extreme case $\beta_0=-5$ (cf.\ Fig.~\ref{spanSTb0}, top
panel). An analogous conclusion holds for collapse of ST polytropes,
cf.\ Sec.~\ref{indata}.
The reason for this absence of spontaneous scalarisation in 
these 
models
lies in the insufficient compactness of 
their protoneutron stars.
At the end of our simulations, the protoneutron star
  in model WH12  has a compactness of
$m/r \sim 0.05$ (at $r = 10$ km), 
  significantly lower than the threshold of $\sim 0.2$ at which multiple
families of stationary solutions appear \cite{2004AIPC..736...35E}.
\end{itemize}
The final compactness reached by NS remnants is naturally model
dependent and the microphysics implemented in our analysis is greatly
simplified by the use of the hybrid EOS. The possibility of triggering
spontaneous scalarisation in stellar core collapse forming NSs (as
opposed to BHs) clearly requires further exploration with more
realistic finite-temperature EOS, which is left to future
work. Dissociation of accreting heavy nuclei at the shock and neutrino
cooling act indeed in the direction of lowering the effective
adiabatic index in the postshock region, thus facilitating a more
rapid increase in 
the protoneutron star's mass and
compactness \cite{2016ApJ...820...76R}. We 
probe
this expectation within our current framework by
evolving the WH12 model with an adiabatic index $\Gamma_{\rm th}$
artificially lowered to 
$1.25$. 
With such a low value of $\Gamma_{\rm th}$,
  the shock stalls at a small radius and material accreting through
  the shock quickly settles onto the protoneutron star, driving up its
  mass and compactness. At the end of our simulation, at
  $0.7\,\mathrm{s}$, the protoneutron star in this model has reached a compactness of 
  $\sim0.18$ and is spontaneously scalarised. 
      Configurations with non-trivial scalar-field
profiles are energetically favoured over their weak-field counterparts
and the dynamical evolutions naturally settle there. The GW strain
$\sqrt{S_h(f)}$ increases by roughly two orders of magnitude when
compared to runs performed using the more realistic value $\Gamma_{\rm
  th}=1.35$.
Galactic signals from spontaneously scalarised NSs, if formed in core
collapse, will 
likely be detectable
by Advanced LIGO even beyond the Cassini bound $\alpha_0=3\times
10^{-3}$. Given its observational potential, this topic definitely
merits further investigation with more realistic microphysics.

\section{Conclusions}
\label{conclusions}
This paper presents an extension of the open-source code
{\sc gr1d} \cite{2010CQGra..27k4103O} to ST theories of gravity.
The required additions to {\sc gr1d} can be summarised as follows:
\begin{enumerate}
\item generalisation of the flux and source terms in the high-resolution
shock capturing scheme according to (\ref{eq:fD})--(\ref{sourcetau})
as well as the constraint equations (\ref{eq:Phir}), (\ref{eq:mr})
for the metric components;
\item implementation of the evolution of the scalar field according to
Eqs.~(\ref{eq:varphit})--(\ref{eq:psit2}) using standard finite differencing;
\item outgoing radiation boundary condition for the scalar field
(\ref{eq:OBCpsi}), (\ref{eq:OBCeta}).
\end{enumerate}
The scalar field furthermore introduces a new radiative degree of
freedom propagating at the speed of light, which requires a
smaller numerical timestep.
    All presented time evolutions start from one of
two types of initial data, (i) polytropic models obtained by solving
the time independent limit of the evolution equations and (ii) more
realistic pre-SN models from zero-age main-sequence stars of masses
$12$ and $40 M_\odot$ \cite{2007PhR...442..269W}.

In this framework, we have simulated a large number of collapse
scenarios which are characterised by five parameters: the linear and
quadratic coefficients $\alpha_0$ and $\beta_0$ determining the
coupling function of the ST theory and the adiabatic indices
$\Gamma_1$, $\Gamma_2$ and $\Gamma_{\rm th}$, characterising the
phenomenological hybrid EOS used in the time
evolution. We summarise our main observations as follows.
\begin{itemize}

  \item The most prominent GW signals are detected from the collapse
    of 
progenitor stars that
      form BHs after a protoneutron star phase
      (such as the $M_{\rm
      ZAMS}=40M_{\odot}$ model of \cite{2007PhR...442..269W}), as
    opposed to collapse events forming 
      long-term
      stable
      NSs. The collapse of protoneutron stars to BHs is the
    most promising dynamical feature for monopole gravitational
    radiation in the context of ST theories.

  \item The dynamical features in the matter fields (density, mass function,
        pressure) resemble closely those obtained in the general
        relativistic limit $\alpha_0=\beta_0=0$. In other
        words, the effect of the scalar field on the matter dynamics
        is weak.

  \item The opposite is not true. The scalar radiation or
        GW {\em breathing mode} is sensitive to
        the specifics of the 
       collapse dynamics
        as well as the choice of ST parameters $\alpha_0$ and $\beta_0$.
        The observed dependencies are of the kind one would intuitively
        expect. EOS and progenitors giving rise to more compact post-collapse configurations
        result in stronger radiation and the amplitude of the scalar wave
        sensitively increases with $\alpha_0$ with approximately
        a quadratic dependence.

  \item The ST parameter $\beta_0$ is known to generate strongly
    non-linear effects in the scalar field for $\beta_0\lesssim
    -4.35$, the so-called {\em spontaneous scalarisation}
    \cite{1993PhRvL..70.2220D,1996PhRvD..54.1474D}.  For
     progenitors collapsing to BHs
   after a protoneutron star phase, transition of
    the central core to a spontaneously scalarised configuration
    before BH formation further enhances the outgoing GW signal. For
    progenitors forming NSs but
    not BHs, we 
    do not find
      spontaneously scalarised configurations for physically
    plausible values of the adiabatic indices in our hybrid EOS. We
    attribute this to the stellar compactness achieved in those
    collapse scenarios being insufficient to trigger spontaneous
    scalarisation. This observation may be an artefact of our
    simplistic treatment of microphysics in 
our      simulations 

  \item We have extracted waveforms from a large set of simulations
        and compared their amplitude for
        the case of a fiducial distance $D=10~{\rm kpc}$ with the
        sensitivity curves of Advanced LIGO and the Einstein
        Telescope. Given the present constraints from the Cassini probe,
        $\alpha_0\lesssim 3\times  10^{-3}$, scalar radiation may be
        marginally detectable from galactic sources. This offers the
        possibility of providing constraints on ST theory with
        GW observations in case of a favourable event occurring
        in the Milky Way. 
        Considerable power is emitted at low frequencies $f\lesssim 10$ Hz, thus making core 	collapse in ST theory ideal sources for future experiments such as DECIGO \cite{2011CQGra..28i4011K}.
        
    \end{itemize}

    The impact of
     more realistic microphysic,
   as for example
nuclear dissociation 
at the shock
and neutrino cooling, on the compactness of the core and, thus, its
degree of scalarisation, represents one key extension left for future
work. Our analysis has shown that the massive increase in wave
amplitudes due to spontaneous scalarisation 
and BH  formation
has the potential to drastically increase the range for
detection.  The $\mathcal{O}(10^3)$ waveforms generated for this work
were completed in less than a week using $\mathcal{O}(10^2)$
CPU cores simultaneously. A moderate increase in the
computational resources will make simulations using tabulated
finite-temperature EOS feasible. Since the matter fields' dynamics is very similar to the GR case, one may perhaps take advantage of existing GR simulations and simulate the scalar field evolution using such GR results as backgrounds. A further numerical improvement may consist in computing (perhaps iteratively) approximate initial conditions for the scalar field from existing pre-collapse stellar models (such as WH12 and WH40 used in this paper), in order to reduce the brief unphysical transient in the GW signal.

Aside from the treatment of the microphysics,
our study offers further scope for extension.
The effects of multiple scalar fields
in ST theories on gravitational collapse
remains largely unknown in spite of some early
studies \cite{1992CQGra...9.2093D} (see \cite{2015CQGra..32t4001H}
for an analysis of  static NS solutions in this framework), but
represents a relatively minor addition to our code. The same holds
for ST theories with non-vanishing potential, as for example massive
fields \cite{2012PhRvD..85f4041A,2016PhRvD..93f4005R}.

As GW physics and astronomy are ushering in a new era,
the community will be offered a wealth of unprecedented
opportunities to observationally test generalisations of GR.
Stellar collapse clearly offers a vast potential for
such fundamental tests.

\section*{Acknowledgments}
We thank
Chris Moore,
Jerome Novak, Evan O'Connor, Norbert Wex, Paulo Freire
and
Carlos Sopuerta
for fruitful discussions. D.G. is supported by the UK STFC and the
Isaac Newton Studentship of the University of Cambridge.  U.S. is
supported by the H2020 ERC Consolidator Grant ``Matter and
strong-field gravity: New frontiers in Einstein's theory'' grant
agreement No. MaGRaTh--646597, the European Union's Horizon 2020 research and innovation programme under the Marie Skludowska-Curie grant agreement 690904, the STFC Consolidator Grant No. ST/L000636/1,
the SDSC Comet and TACC Stampede clusters through NSF-XSEDE Award
Nos.~TG-PHY090003 and TG-PHY100033, the Cambridge High Performance
Computing Service Supercomputer Darwin using Strategic Research
Infrastructure Funding from the HEFCE and the STFC, and DiRAC's Cosmos
Shared Memory system through BIS Grant No.~ST/J005673/1 and STFC Grant
Nos.~ST/H008586/1, ST/K00333X/1. C.D.O. is partially supported by NSF
under award Nos.~CAREER PHY-1151197, and PHY-1404569, and by the
International Research Unit of Advanced Future Studies, Kyoto
University.  Figures were generated using the \textsc{python}-based
\textsc{matplotlib} package \cite{2007CSE.....9...90H}. This article
has been assigned Yukawa Institute report number YITP-16-14.

\section*{References}

\bibliographystyle{iopart-num_davide}
\bibliography{corecollapseST}

\end{document}